\documentclass[conference]{IEEEtran}
\IEEEoverridecommandlockouts

\usepackage{cite}
\usepackage{amsmath,amssymb,amsfonts}
\usepackage{algorithmic}
\usepackage{graphicx}
\usepackage{textcomp}
\usepackage{xcolor}
\usepackage{algorithm}
\usepackage{array}
\usepackage{ulem} 
\usepackage[caption=false,font=normalsize,labelfont=sf,textfont=sf]{subfig}
\usepackage{textcomp}
\usepackage{stfloats}
\usepackage{url}
\usepackage{verbatim}
\usepackage{epstopdf} 
\usepackage{amssymb}
\def\BibTeX{{\rm B\kern-.05em{\sc i\kern-.025em b}\kern-.08em
    T\kern-.1667em\lower.7ex\hbox{E}\kern-.125emX}}

\begin{document}

\title{Blockage-aware Hierarchical Codebook Design for RIS-Assisted Movable Antenna Systems \\
\thanks{Y. Zhang is with the Department of Electronic and Electrical Engineering, Trinity College Dublin (e-mail: zhangy42@tcd.ie). This work was supported in part by the Chinese Scholarship Council. This work was also supported in part by Science Foundation Ireland under Grant 13/RC/2077\_P2, and by the EU MSCA Project ``COALESCE'' under Grant Number 101130739.}

}

\author{\IEEEauthorblockN{Yan Zhang}
\IEEEauthorblockA{\textit{dept. Electronic and Electrical Engineering} \\
\textit{Trinity College Dublin}\\
Dublin, Ireland \\
zhangy42@tcd.ie}
\and
\IEEEauthorblockN{Indrakshi Dey}
\IEEEauthorblockA{\textit{dept. Computing and Mathematics} \\
\textit{South East Technological University}\\
Waterford, Ireland \\
Indrakshi.Dey@waltoninstitute.ie}
\and
\IEEEauthorblockN{Nicola Marchetti}
\IEEEauthorblockA{\textit{dept. Electronic and Electrical Engineering} \\
\textit{Trinity College Dublin}\\
Dublin, Ireland \\
nicola.marchetti@tcd.ie}
}

\maketitle

\begin{abstract}
In this paper, we propose a novel blockage-aware hierarchical beamforming framework for movable antenna (MA) systems operating at millimeter-wave (mm-Wave) frequencies. While existing works on MA systems have demonstrated performance gains over conventional systems, they often neglect the design of specialized codebooks to leverage MA's unique capabilities and address the challenges of increased energy consumption and latency inherent to MA systems. To address these aspects, we first integrate blockage detection into the codebook design process based on the Gerchberg-Saxton (GS) algorithm, significantly reducing inefficiencies due to beam evaluations done in blocked directions. Then, we use a two-stage approach to reduce the complexity of the joint beamforming and Reconfigurable Intelligent Surfaces (RIS) optimization problem. The simulations demonstrate that the proposed adaptive codebook successfully improves the Energy Efficiency (EE) and reduces the beam training overhead, substantially boosting the practical deployment potential of RIS-assisted MA systems in future wireless networks. 
\end{abstract}


\section{Introduction}
\IEEEPARstart{A}{s} fifth-generation (5G) systems approach maturity, sixth-generation (6G) research is targeting terabit per second throughput and sub-millisecond latency through terahertz (THz) frequencies \cite{karacora2025robust}. Massive multiple-input multiple-output (MIMO) systems also represent an important technology for 6G, leveraging large antenna arrays to achieve substantial spatial multiplexing gains. However, THz bands suffer from severe path loss and environmental blockages, while contemporary MIMO technologies rely on fixed-position antennas that inherently constrain their capacity to leverage spatial channel variations or respond adaptively to dynamic propagation phenomena \cite{wang2024tutorial}.

Movable antenna (MA) technology has emerged as a novel paradigm to improve communication reliability in high-frequency and dynamic wireless environments. By enabling mechanical repositioning of antennas within a confined spatial region, MA-assisted systems can dynamically achieve more favorable channel conditions to better serve User Equipments (UEs) \cite{zhu2023modeling}. The ability to exploit spatial diversity through antenna mobility holds considerable promise, especially in millimeter-wave (mmWave) bands, where line-of-sight (LOS) propagation is often obstructed by environmental obstacles. This advancing feature of MA can serve as a valuable complement to reconfigurable intelligent surfaces (RIS), which have limited capability to fully exploit the available spatial degrees of freedom. The synergistic integration of MA and RIS technologies presents unprecedented opportunities for next-generation wireless networks, offering enhanced spatial reconfigurability beyond what either technology can achieve independently. 

Despite these advantages, MA systems face notable challenges in practical deployments that limit their widespread adoption. The frequent repositioning of antennas introduces substantial management burden and energy consumption. As highlighted in \cite{shao2025tutorial}, continuous antenna adjustments required to adapt to channel fluctuations result in high mechanical control costs that can quickly deplete battery resources. Moreover, mechanical movements contribute to wear-and-tear, demanding regular calibration and maintenance, especially in delay-sensitive applications like vehicular or satellite communication \cite{ning2024movable}. These operational constraints raise fundamental concerns regarding the long-term scalability, reliability, and energy sustainability of MA systems, particularly in resource-constrained scenarios. 

Additionally, real-world wireless networks are often affected by dynamic and dense blockages that conventional MA designs fail to adequately address. Studies such as \cite{bai2013coverage} illustrate the severe impact of environmental blockages on signal degradation, including increased path loss, delay spread, and coverage holes. Although MA systems are more suitable for delay-tolerant applications such as Internet of Things (IoT) \cite{shao2025tutorial}, these scenarios are typically energy-constrained, and embedding MA in such devices can further complicate energy management with significant repercussions on battery life \cite{palitharathna2023neural}. The situation is exacerbated in dense urban environments and indoor scenarios, where human bodies, furniture, and structural elements create time-varying blockage patterns that dynamically alter channel conditions. Furthermore, MA positioning itself can introduce additional blockages when antennas move to certain positions, creating a dual-blockage problem that existing designs overlook. These considerations emphasize the urgent need for efficient, blockage-aware MA schemes that minimize energy expenditure while maintaining robust beamforming performance under realistic propagation conditions. 

Therefore, critical gaps remain unresolved in existing MA research. While exhaustive beam search could provide optimal performance for MA systems, its prohibitive beam training overhead exacerbates time resource consumption \cite{lu2023hierarchical}. Current hierarchical codebook designs often assume idealized conditions and neglect dual blockage sources — environmental obstacles and MA-induced blockages in particular positions — resulting in substantial energy resource waste on beam training in blocked directions \cite{hwang2024movable}, which is particularly detrimental given MA systems' already elevated resource requirements, imposing fundamental limitations that significantly degrade MA system performance. 

Considering the aforementioned concerns, we study the beam training problem for MA-assisted base stations operating with RIS support under realistic blockage conditions. Our objective is to develop a blockage-aware hierarchical codebook framework that minimizes beam training overhead and energy waste by intelligently avoiding blocked directions. We propose a GS-based hierarchical codebook specifically tailored for MA systems and employ a two-stage optimization approach: first, RIS phase shifts are optimized using statistical Channel State Information (CSI); second, efficient beam training is performed in the RIS-enhanced channel. Through extensive simulations in dynamically changing blockage-prone environments, we demonstrate that our framework significantly improves beamforming performance while reducing resource consumption compared to conventional approaches. 

\section{Related works}
\subsection{Movable Antenna Systems }
The concept of movable antennas has gained significant attention as a promising technology to enhance wireless communication performance through dynamic spatial diversity exploitation. The foundational tutorial in \cite{zhu2023movable, zhu2024movable} provided a comprehensive overview of MA technologies, highlighting their opportunities and challenges in exploiting spatial channel variations within confined regions. Robust position optimization considering imperfect CSI with norm-bounded and statistical errors has been investigated \cite{ma2025robust}. Statistical CSI-based approaches \cite{yan2025movable, chen2023joint} reduce antenna movement overhead by optimizing positions over large timescales. Derivative-free optimization methods \cite{zeng2024csi, zeng2025derivative} using zeroth-order gradient approximation enable CSI-free position optimization. Fundamental capacity characterization \cite{ma2023mimo} and wireless sensing applications \cite{ma2024movable} have demonstrated significant performance improvements over fixed arrays. However, these works predominantly assume idealized propagation conditions without considering the dual impact of environmental blockages and MA position-dependent obstructions. Furthermore, existing MA optimization methods focus on maximizing instantaneous performance metrics such as signal-to-interference-plus-noise ratio and spectral efficiency without addressing the energy waste incurred by training beams in blocked directions, which is particularly critical for energy-constrained IoT deployments.

\subsection{Beam Training and Codebook Design }
Efficient beam training remains a critical challenge in mmWave systems due to large antenna arrays and narrow beams required for adequate coverage. Exhaustive beam search, while theoretically optimal, requires testing all beam-position combinations, resulting in training overhead proportional to the product of codebook size and candidate MA positions. This approach becomes impractical for large arrays, often requiring hundreds of training slots and exceeding channel coherence time \cite{brilhante2023literature}. To mitigate this, hierarchical framework designs have emerged as standard approaches to reduce training overhead compared to exhaustive search methods \cite{eom2025hierarchical}, \cite{xiao2016hierarchical}, \cite{ wang2009beam}. The GS orthogonalization process has been adopted for generating beamforming codebooks due to its ability to create structured beam patterns with reduced inter-beam interference \cite{matsumura2011orthogonal}. Advanced hierarchical schemes \cite{wang2023hierarchical, shi2023spatial} have proposed dynamic codebooks that update according to estimated multi-path components and spatial-chirp patterns for extremely large-scale MIMO systems. Adaptive beam alignment techniques \cite{liu2022adaptive} with variable training time based on SNR detection improve success rates while reducing overhead. Despite these advances, existing hierarchical codebook designs assume uniform spatial coverage and do not incorporate blockage awareness into the codebook structure itself. These methods waste significant training resources by exhaustively searching directions that are known to be blocked or severely attenuated. Moreover, conventional codebooks are not specifically tailored for MA systems where antenna positioning creates additional position-dependent blockage constraints that vary across the movement region.

\subsection{Blockage Modeling in Wireless Networks}
Understanding and mitigating blockage effects has become increasingly important in mmWave and beyond-5G systems. Analysis for urban cellular networks \cite{6840343, ramirez2024observations} quantifies blockage effects on coverage and capacity. Foundational studies using stochastic geometry \cite{alyosef2022survey} characterized how environmental obstacles and human-body blockages cause signal degradation through increased path loss and delay spread. Extensive measurements \cite{azpilicueta2020fifth, gapeyenko2016analysis} demonstrated that static obstructions like concrete walls can reduce signals by 40-60 dB, while mobile obstacles such as pedestrians cause 20-35 dB signal loss. Research on blockage tolerance \cite{liu2021blockage} investigated novel blockage models combined with stochastic geometry for mmWave backhaul networks. However, these blockage models primarily focus on characterizing and predicting blockage events rather than incorporating blockage awareness directly into the beam training codebook design. Existing works treat blockage as a channel impairment to be mitigated through prediction or avoidance, but do not fundamentally redesign the beam training process to intelligently skip blocked directions. Furthermore, the dual-blockage problem in MA systems — where both environmental obstacles and MA positioning create coupled blockage effects—remains unexplored in the literature. 

\subsection{Joint RIS and MA Optimization }
While RIS and MA technologies have been studied extensively in isolation, their joint optimization remains a developing research area. RIS has demonstrated effectiveness in overcoming blockages by creating alternative propagation paths \cite{wu2025intelligent}, but faces limitations in fully exploiting spatial degrees of freedom. Recent work \cite{geng2025joint}, \cite{yu2025achievable}, \cite{yang2025robust} investigated joint beamforming and antenna position optimization for MA-enabled RIS systems using product Riemannian manifold optimization, alternating optimization frameworks, and fractional programming-based approaches. Existing approaches face significant computational complexity in joint beamforming and phase shift optimization that scales poorly with system size. The development of practical, low-complexity frameworks for MA-RIS coexistence, particularly under realistic blockage conditions, represents a critical gap in current literature. 

\subsection{\textbf{Energy Efficiency Considerations}}
Energy management is crucial for MA systems, especially in IoT deployments. Research \cite{callebaut2021art}, \cite{rzig2025energy}, \cite{zhao2024joint} has addressed energy-efficient optimization and power management strategies. The mechanical movements in MA systems contribute to increased power consumption, which is particularly challenging for battery-constrained IoT applications \cite{tang2021battery}. Studies on dynamic metasurfaces \cite{di2020smart} have highlighted similar trade-offs between adaptability and energy consumption in programmable electromagnetic environments. Recent advances in energy harvesting \cite{mukherjee2022leveraging} have explored 5G-powered wireless grids for IoT devices, while \cite{perera2013context} investigates context-aware computing for efficient energy utilization in IoT networks. Solar-powered and RF energy harvesting techniques \cite{shaikh2016energy} offer promising solutions for extending battery life in MA-enabled systems. However, the additional energy requirements imposed by MA positioning mechanisms — including actuator control, sensor feedback, and real-time computation — remain a significant concern for practical deployments. The energy overhead of continuous beam steering and environmental adaptation in MA systems can offset potential communication gains, necessitating careful system-level design and intelligent duty-cycling strategies \cite{bjornson2017massive}. 

In summary, critical gaps remain unresolved in existing MA research. Current approaches assume idealized propagation conditions and design hierarchical codebooks without blockage awareness, resulting in substantial energy waste by exhaustively training beams in blocked or severely attenuated directions. This problem is particularly acute in MA systems due to dual blockage sources—environmental obstacles and position-dependent MA-induced obstructions—that create coupled blockage effects unexplored in existing literature. Moreover, joint RIS-MA optimization methods face prohibitive computational complexity while the additional energy overhead from MA positioning mechanisms threatens to offset potential communication gains in practical deployments. To address these fundamental limitations, we propose a blockage-aware hierarchical codebook framework that explicitly integrates blockage characteristics into the beam training structure. By intelligently skipping blocked directions and leveraging RIS for blockage mitigation, our approach simultaneously reduces time, energy, and computational resource consumption while maintaining search optimality in realistic MA deployments. 

\section{Contributions}
We study the joint beamforming and RIS configuration design in practical RIS-assisted MA-enabled MU-MIMO communication systems under realistic blockage conditions. Unlike conventional approaches, our framework explicitly accounts for dual blockage sources to minimize wasted beam training resources and improve energy efficiency. The main contributions of this paper are as follows: 
\begin{itemize}
    \item \textit{A GS-based hierarchical codebook tailored for MA}:  We design a novel hierarchical codebook specifically optimized for MA-assisted base stations operating in blockage-prone environments. By leveraging the GS algorithm at each layer of the hierarchical codebook, our approach intelligently rotates beams toward target steering directions while avoiding blocked spatial regions, enabling efficient space-division multiple access with reduced beam training overhead compared to exhaustive search methods.
    \item     \textit{A two-stage procedure for beamforming and RIS configuration optimization}: To address the computational complexity challenge in joint MA positioning and RIS configuration, we propose a practical two-stage approach. In the first stage, RIS phase shifts are optimized using statistical CSI without requiring instantaneous channel knowledge, significantly reducing feedback overhead. In the second stage, with the pre-configured RIS enhancing the channel conditions, our blockage-aware hierarchical codebook efficiently performs beam training to identify optimal MA positions and beamforming vectors. 
    \item \textit{Realistic evaluation under dynamic blockage conditions}: We conduct extensive simulations in realistic, dynamically changing environments where signal blockages occur frequently due to moving obstacles and time-varying channel conditions — an often-overlooked yet critical aspect in practical deployment scenarios. Our evaluation demonstrates significant performance gains in spectral efficiency, energy consumption, and beam training overhead compared to conventional codebook-based approaches that neglect blockage awareness. 
\end{itemize}

\section{System and Channel Model}
\begin{figure}[!t]
\centering
\includegraphics[width=0.99\linewidth, trim=0 0 0 10, clip]{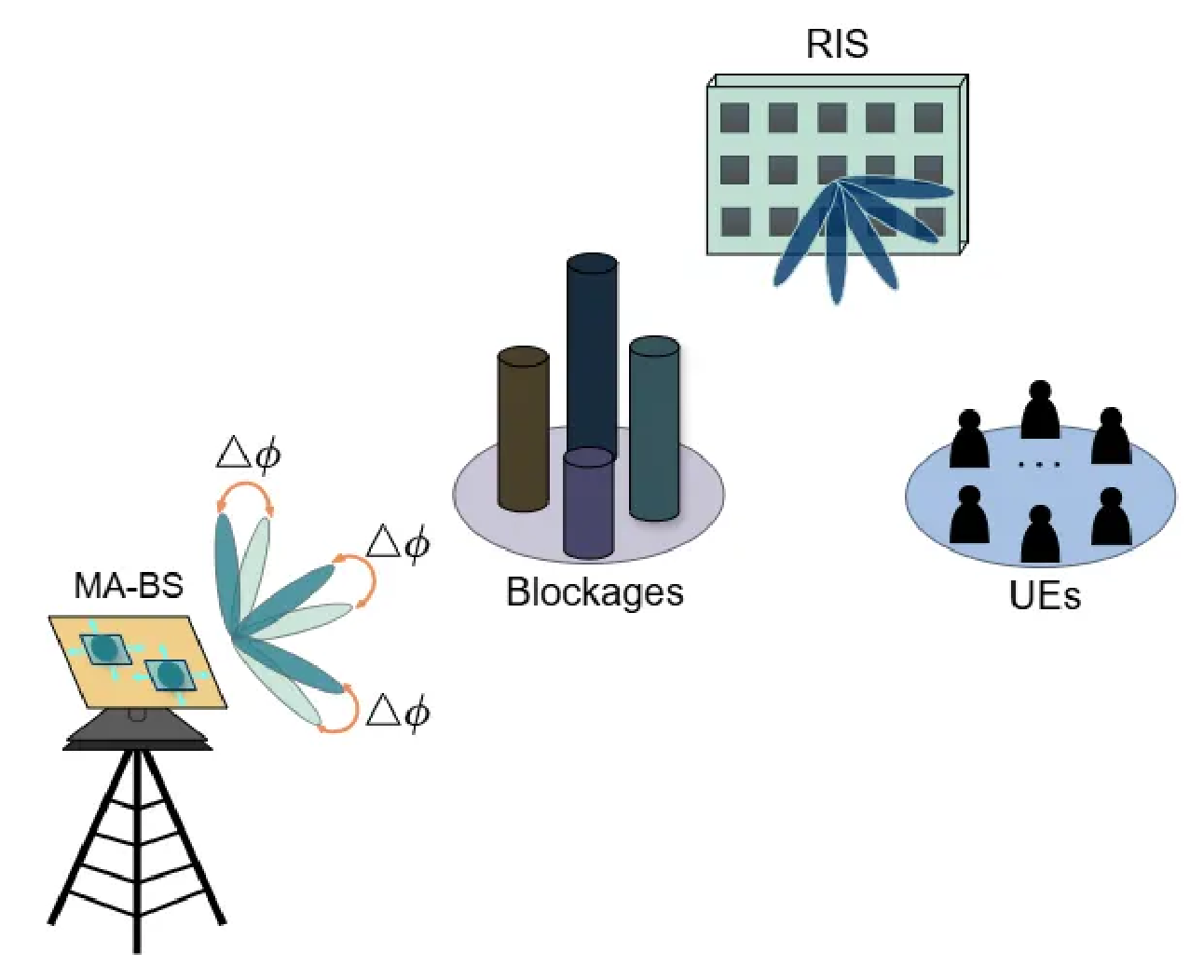}
\caption{System model of MA-BS and RIS-assisted communication illustrating signal obstructions and beam rotation, where $\triangle \phi$ is the beam rotation angle. }
\label{SysModel}
\end{figure}

We consider a downlink scenario where a BS is equipped with $M$ movable antennas, referred to as a Movable-Antenna-Enhanced Base Station (MA-BS). An RIS with $N$ elements is deployed to assist communication, and $K$ single-antenna UEs are served as illustrated in Fig. \ref{SysModel}. The channel between the MA-BS and RIS is subject to blockage, creating a challenge for traditional codebook-based beamforming schemes.
\subsection{Channel model}
The channel between the MA-BS and the RIS is denoted by $\mathbf{H}_{\text{BR}} \in \mathbb{C}^{N \times M}$, while the channel between the RIS and the $k$-th UE is represented by $\mathbf{H}_{\text{RU},k} \in \mathbb{C}^{N \times 1}$, $k \in \mathcal{K} \triangleq  \{ 0, \dots, K-1\} $. The end-to-end channel from the MA-BS to the $k$-th UE through the RIS can be expressed as:
\begin{equation}
    \mathbf{H}_{k} =  \mathbf{H}^H_{\text{RU},k} \boldsymbol\varphi \mathbf{H}_{\text{BR}} \in \mathbb{C}^{1 \times M }
\end{equation}
where $\boldsymbol\varphi=[\varphi_{1},\dots,\varphi_{N}]^T \in \mathbb{C} ^{N \times 1}$ is the RIS phase shift vector with $|\varphi_n| = 1$ for $n = 1, 2, \ldots, N$.
 
The position of $m$-th MA can be represented by Cartesian coordinates $\textbf{\textit{b}}^{2D}_m = [m_x,m_y]^T \in \mathcal{C}$ where $\mathcal{C}$ is the given two-dimensional (2D) region within which the MAs can move freely. Since the MAs are mounted on the BS, whose height is fixed, the actual 3D position of the MA in the global coordinate system can be expressed as $\textbf{\textit{b}}^{3D}_m = [m_x,m_y,m_z]^T \in \mathcal{C}$  where $m_z$ is the BS height. Let $\mathbf{b} \triangleq  [\textbf{\textit{b}}^{3D}_1, \textbf{\textit{b}}^{3D}_2, \cdots, \textbf{\textit{b}}^{3D}_M ]^T \in \mathbb{R}^{3 \times M}$ denote the concatenated vector of all MA positions. Then, we adopt the geometric channel model for the MA-BS to RIS channel $\mathbf{H}_{\text{BR}}$ and RIS to UE channel $\mathbf{H}_{\text{RU},k} $, which is widely used in mmWave communications \cite{wang2020channel}, where $\mathbf{H}_{\text{BR}}$ can be expressed as:
\begin{align}
\mathbf{H}_{\text{BR}} 
&= \sum_{l=1}^{L_g} \rho_{br}^l \mathbf{a}_R^r(\theta_{BR}^l) 
    \left( \mathbf{a}_B(\mathbf{b}, \phi_{BR}^l) \right)^{H} \notag \\
&= \mathbf{A}_R^r(\theta_{BR}) \, \mathrm{diag}(\mathbf{\rho}_{br}) \, 
    \mathbf{A}_B(\mathbf{b}, \phi_{BR})^{H} \in \mathbb{C}^{N \times M}
\label{eq2}
\end{align}
where $L_g$ is the number of propagation paths, $\rho^l_{br} $ is the complex path loss of the $l$-th path. $\mathbf{a}^r_R(\theta^l_{BR}) \in \mathbb{C} ^{N \times 1}$ and $\mathbf{a}_B(\mathbf{b},\phi^l_{BR}) = \frac{1}{M}\left [e^{-j\frac{2 \pi}{\lambda}(\textbf{\textit{b}}^{3D}_1-\textbf{\textit{b}}_0)},\cdots,e^{-j\frac{2 \pi}{\lambda}(\textbf{\textit{b}}^{3D}_M-\textbf{\textit{b}}_0)}  \right ]\in \mathbb{C} ^{M \times 1}$ respectively represent the array response vector at RIS and MA-BS, where $\textbf{\textit{b}}_0$ is the reference point. $\theta^l_{BR}$ and $\phi^l_{BR}$ respectively represent the angle of arrival at RIS and the Angle of Departure (AoD) from the MA-BS of the $l$-th path. 
$\mathbf{A}_R^r(\theta_{BR})= [\mathbf{a}_R(\theta^1_{BR}),\cdots,\mathbf{a}_R(\theta^{L_g}_{BR})] \in \mathbb{C} ^{N\times L_g}$, $\mathbf{A}_B(\mathbf{b},\phi_{BR})= [\mathbf{a}_B(\textbf{\textit{b}}^{3D}_1,\phi^1_{BR}),\cdots,\mathbf{a}_B(\textbf{\textit{b}}^{3D}_M,\phi^{L_g}_{BR})]\in \mathbb{C} ^{M \times L_g} $, and $\mathbf{\rho}_{br} = [\rho_{br}^1, \cdots, \rho_{br}^{L_g}]^T$.
Similarly, the RIS-UE channel $\mathbf{H}_{\text{RU},k}$ can be expressed as:
\begin{equation}
    \mathbf{H}_{\text{RU},k} =  \sum_{l=1}^{L_b}  \rho^l_{ru}  \mathbf{a}^t_R(\phi_{RU}) \in \mathbb{C}^{N \times 1 }
\end{equation}
where $L_b$ is the number of paths between the RIS and the UE, $\rho^l_{ru} $ is the complex gain of the $l$-th path, and $\phi_{RU}$ is the AoD from the RIS, $\mathbf{a}^t_R(\phi_{RU})$ is the transmit steering vector from RIS. The received signal of the $k$-th UE is expressed as
\begin{align}    \label{eq}
y_{k} =  \sum_{k=1}^K \sqrt{p} \mathbf{H}_{k} \mathbf{w} s_{k} + n_{k}
\end{align}
where $\mathbf{w} \in \mathbb{C} ^{M \times 1}$ is the beamforming vector at MA-BS, $s_{k} $ is the training symbol of $k$-th UE which is designed as $s_{k} = 1 $ for the sake of exposition simplicity, $n_{k} \sim \mathcal{CN} \left \{ 0, \sigma^2 \right \}  $ is the additive white Gaussian noise (AWGN) at the UE, $p$ is the pilot transmit power. Then the received signal to noise ratio (SNR) at UE $k$ is
\begin{equation}
    \gamma_k = \frac{p \left | \mathbf{h}_{k}^H \mathbf{w}   \right | ^2 }{\sigma^2}
\end{equation}
So that the achievable transmit rate of UE can be expressed as
\begin{equation}
   R = \log_2(1+ \gamma_k) = \log_2\left(1+  \frac{p \left | \mathbf{H}_{k}^H \mathbf{w}   \right | ^2 }{\sigma^2}\right) 
\end{equation}
\subsection{Path Loss}
We follow \cite{galiotto2017effect} to consider a realistic path loss model encompassing blockages for the MA-BS to RIS link as: 
\begin{equation}\label{e1}
\rho^l_{br}(d) = \begin{cases}
K_L d^{-\beta_L} & \text{with probability } p_L(d) \\
K_{NL} d^{-\beta_{NL}} & \text{with probability } 1 - p_L(d)
\end{cases}
\end{equation}
where $\beta_L$ and $\beta_{NL}$ are the path-loss exponents for LOS and NLOS propagation, respectively; $K_L$ and $ K_{NL}$ are the signal attenuation at distance $ d = 1 $m for LOS and NLOS propagation, respectively; $p_L(d)$ is the probability of having LOS as a function of the distance $d$.
To ensure that our formulation and the outcomes of our study are general and not limited to a specific LOS probability pattern, we consider the LOS probability function as:
\begin{equation}\label{e2}
    p_L(d)  = \mathrm{exp}(-(d/L)^2)
\end{equation}
where $L$ is the parameter that allows (\ref{e2}) to be tuned to match (\ref{e1}).

\subsection{Energy Efficiency Formulation}
We define the EE of the system as the ratio between the sum-rate that the system can achieve and the total power consumed (units: bits/Joule) \cite{guo2015power, wei2025mechanical}

\begin{equation}
    \eta_{EE} = \frac{(T-\tau)C}{E_{total} } =\frac{(T-\tau)C}{\tau P_{M} + (T-\tau)\frac{P_{D}}{\eta_{amp}} } 
\end{equation}
where $C$ is the channel capacity, corresponding to the maximum theoretical transmission rate limit. $\tau$ is time spent on MA's mechanical move, $T$ is the given time slot. $E_{total}$ is the total power consumption of MA-BS, which comprises the energy radiated for data transmission and that consumed by the MA motion, i.e., 
\begin{equation}
     E_{total} = \tau P_{M} + (T-\tau)\frac{P_{D}}{\eta_{amp}} 
 \end{equation}
where $\eta_{amp}$ is power amplifier efficiency, $P_M$ and $P_D$ are the power consumption due to MA motion and during data transmission, respectively. $P_D$ includes the radiated power used for beam training and the static circuit power consumption, which is given by \cite{guo2015power},
\begin{equation}
    P_D =M P_c + P_s 
\end{equation}
where $P_c$ is the dynamic power consumption proportional to the number of MAs, and $P_s$ is the static circuit power consumption, respectively.

\section{Proposed Two-Stage Method - Codebook Design and Hierarchical Beam Training}
\label{sec:algo} 
Our objective is to design a hierarchical codebook that dynamically adapts to blockages existing between the MA-BS and RIS, while minimizing energy consumption and beam training overhead at MA-BS side. The key idea is to incorporate blockage detection into the codebook design, enabling the system to maintain acceptable data rates and EE under blockage conditions. 
\subsection{Blockage Detection}
We assume the UEs are located in the near field of RIS, so that we do not experience blockage in the RIS-UEs channel. 
\begin{figure}[!t]
\centering
\includegraphics[width=1\linewidth, trim=0 10 10 35, clip]{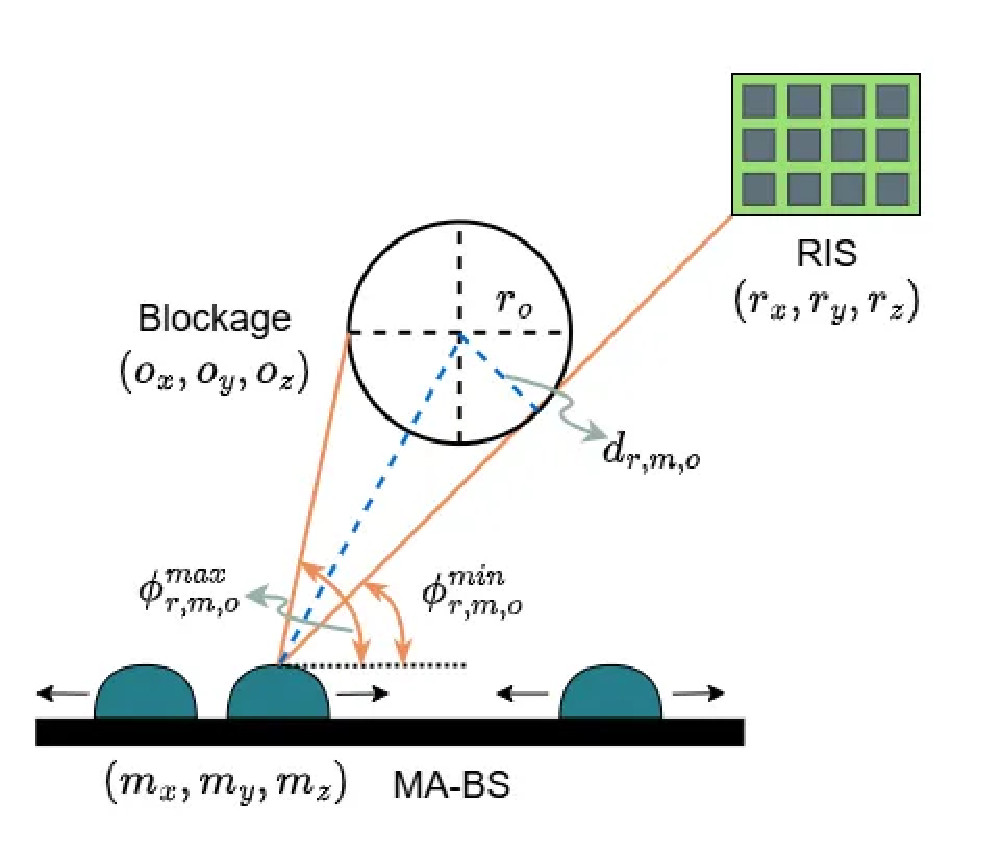}
\caption{Illustration of blockage geometry.}
\label{BlockageIll}
\end{figure}
We suppose the $o$-th blockage (located at $(o_x,o_y,o_z)$) has a circular shape with radius $r_o$ as shown in Fig. \ref{BlockageIll}. The 3D position of $n$-th RIS element is $\textbf{\textit{r}}^{3D}_n = [r_x,r_y,r_z]^T$. The distance $d_{r,m,o}$ between the center of the $o$-th blockage and the direct line between the $m$-th MA and the $n$-th RIS element, is defined as:
\begin{equation}
    d_{r,m,o} =\frac{\left | (r_y -m_y)o_x-(r_x -m_x)o_y +r_xm_y -m_xr_y \right | }{\sqrt{(r_y-m_y)^2+(r_x-m_x)^2}}
\end{equation}
It can be noticed that when $d_{r,m,o} < r_o$, there is no intersection. Then, the blocked angle is defined as:
\begin{equation}
    \phi^{min}_{r,m,o} = \mathrm{tan}^{-1}\left \{ \frac{r_z-m_z}{sq_a }   \right \} 
\end{equation}
\begin{equation}
    \phi^{max}_{r,m,o} = \mathrm{tan}^{-1}\left \{ \frac{o_z}{sq_a-d^2_{r,m,o} -sq_b}   \right \} 
\end{equation}
where $sq_a = \sqrt{(r_x-m_x)^2+(r_y-m_y)^2}$, and $sq_b=\sqrt{r_o^2-d^2_{r,m,o}}$. Hence, the $o$-th blockage detection range would be $\left [  \phi^{min}_{r,m,o} ,  \phi^{max}_{r,m,o} \right ]  $. Then, we can subtract all blocked angular ranges from the total angular coverage:
\begin{equation}
    \phi_{ava} = \phi \in [\phi_{min}, \phi_{max}]\setminus \underset{o}{U}[\phi^{min}_{r,m,o} ,\phi^{max}_{r,m,o} ]. 
\end{equation}
where $\underset{o}{U}[\phi^{min}_{r,m,o} ,\phi^{max}_{r,m,o} ] $ is the collection of all blocked angle ranges. The available angle range $ \phi_{ava}$ becomes the new target range for beam rotation, allowing the MA-BS to redirect its beam toward unobstructed directions.
\subsection{Beam Rotation}
\begin{figure}[!t]
\centering
\includegraphics[width=0.99\linewidth, trim=0 0 0 25, clip]{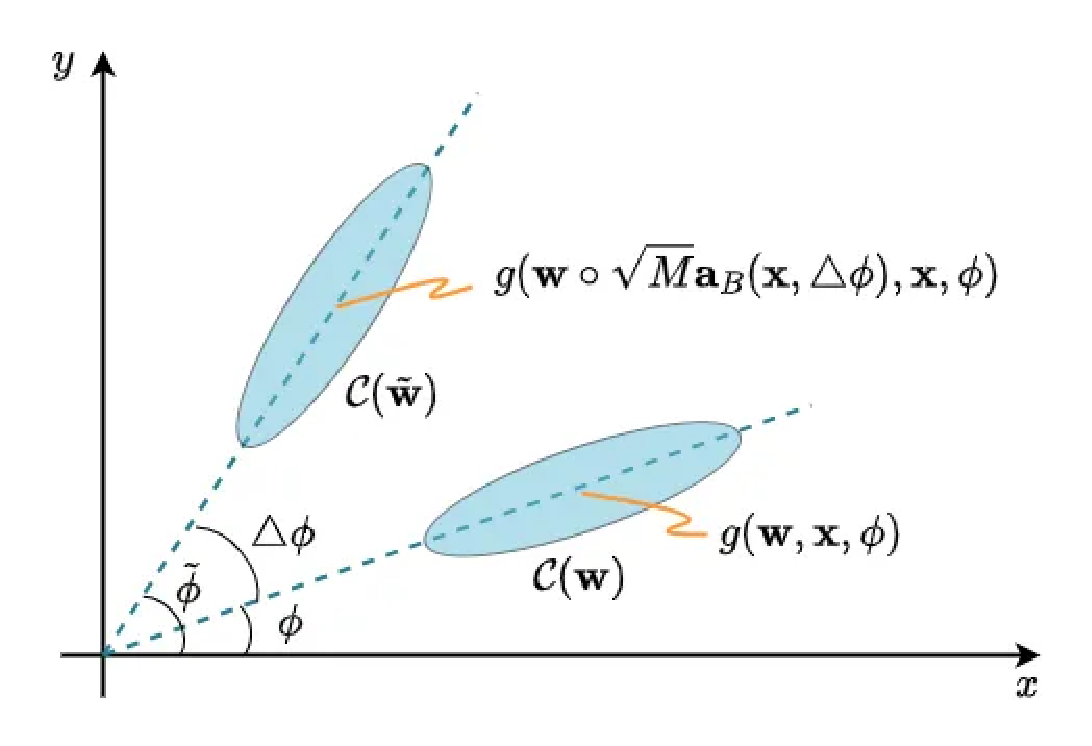}

\caption{Illustration of beam rotation.}
\label{BeamRo}
\end{figure}

Denote $\tilde{\phi}  = \phi \pm \triangle \phi$ as the target rotation angle, where $\triangle \phi$ is the rotation angle, $\phi$ is the blocked angle direction. Let $\mathbf{g}_w$ denote the desired beam pattern of codeword $\mathbf{w}$ for a target steering angle $\tilde{\phi}$, 
\begin{equation}
    \mathbf{g}_w = [g(\mathbf{w} ,\mathbf{b},\tilde{\phi}_1), g(\mathbf{w} ,\mathbf{b},\tilde{\phi}_2)\cdots, g(\mathbf{w} ,\mathbf{b},\tilde{\phi}_M)  ]
\end{equation}
where $g(\mathbf{w} ,\mathbf{b},\tilde{\phi}) = \left | g(\mathbf{w},\mathbf{b},\tilde{\phi}) \right | e^{jf(\mathbf{w} ,\mathbf{b},\tilde{\phi} )}$ is the target beam gain \cite{lu2023hierarchical}, with $f(\mathbf{w} ,\mathbf{b},\tilde{\phi} )$ representing the phase information. As shown in Fig. \ref{BeamRo}, the coverage of the beam pattern also rotates along the target beam direction, thus the coverage rotation can be formulated as \cite{chen2023hierarchical}
\begin{equation}    \label{eq2}
\begin{split}
\mathcal{C}(\tilde{\mathbf{w}})   &= \left \{\phi | g(\mathbf{w} \odot \sqrt{M} \mathbf{a}_B(\mathbf{b}, \triangle \phi) ,\mathbf{b},\phi) > \rho\right \} \\
&= \left \{\phi|g(\mathbf{w},\mathbf{b}, \tilde{\phi} ) > \rho\right \}
\end{split}
\end{equation}
where $\rho$ is the minimum required gain in each layer. 

\subsection{Stage-I RIS Phase-Shift Optimization}
The angular-domain near-field beam training could employ exhaustive search over an angular-domain codebook constructed based on the Discrete Fourier Transform (DFT). The codebook comprises $M$ orthogonal steering vectors uniformly distributed across the angular space as $\mathbf{A}= [\mathbf{a}_B(\textbf{\textit{b}}^{3D}_1,\phi^1_{BR}),\cdots,\mathbf{a}_B(\textbf{\textit{b}}^{3D}_M,\phi^{M}_{BR})] $ where $\phi^m_{BR} = \frac{2 \pi(m-1)}{M}, m=1,2,\cdots,M$. The objective of codeword design is to generate beamforming weights $\mathbf{w}$ such that the resulting beam pattern $\mathbf{A}^H \mathbf{w}$ closely approximates the desired beam pattern $\mathbf{g}_w$ directed toward a specific target angle. Consequently, the theoretical codeword optimization problem for steering angle $\tilde{\phi}$ can be formulated as: 
\begin{equation}\label{e3}
    \mathrm{\min_{\mathbf{w}, f(\mathbf{w} ,\mathbf{b},\tilde{\phi} )}} \left \| \mathbf{A}^H \mathbf{w} -\mathbf{g}_w \right \|^2_2, \\    (\mathrm{P1})
\end{equation}
We adapt the GS framework to address the fundamental problem of determining optimal beam rotation angles in blockage-aware scenarios. Rather than pursuing theoretical ideal patterns like \cite{lu2023hierarchical}, our approach leverages the GS algorithm to solve the beam steering optimization under spatial constraints. Specifically, we modify the conventional codeword design objective in (P1) to incorporate blockage detection results, effectively transforming the phase retrieval problem into a blockage-aware beam rotation optimization. This modification enables our system to dynamically adapt beam patterns based on blockage conditions. 

\begin{figure}[!t]
\centering
\includegraphics[width=0.99\linewidth, trim=0 0 0 25, clip]{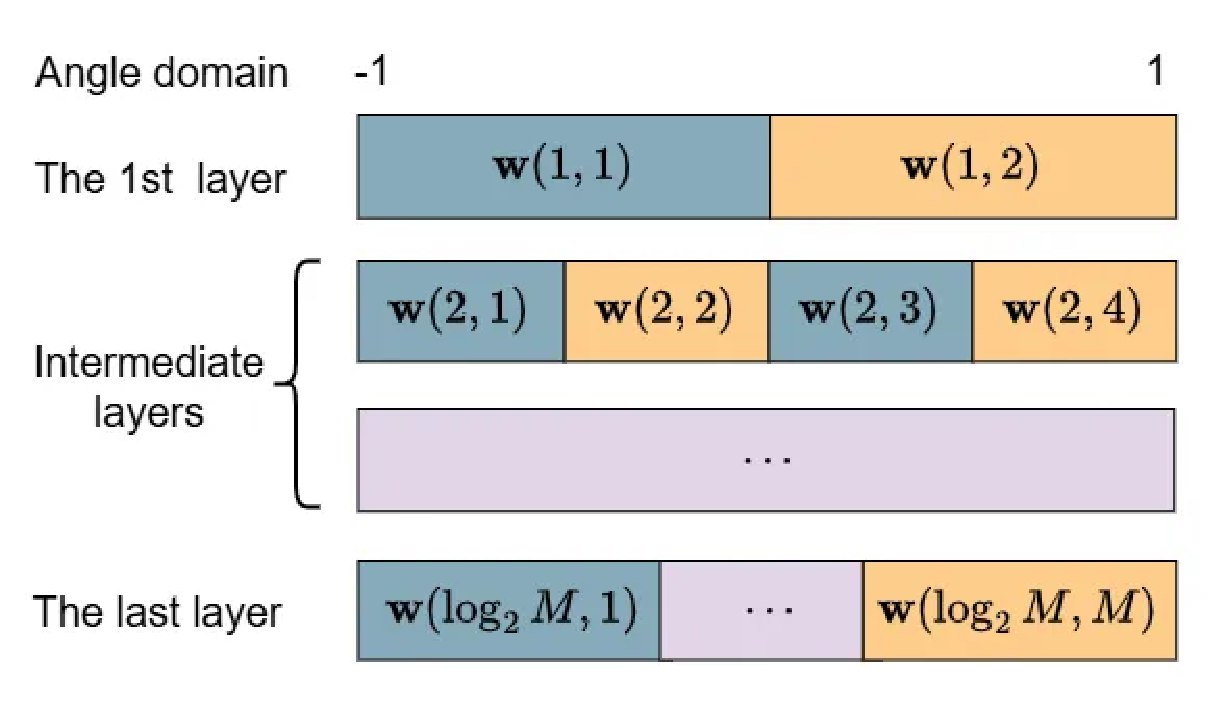}

\caption{Binary-tree structured hierarchical codebook.}
\label{HieCodebook}
\end{figure}

As shown in Fig. \ref{HieCodebook}, the proposed codebook employs a coarse-to-fine binary tree structure that progressively narrows beam coverage across multiple layers. In fact, $\phi^l_{BR} = \mathrm{cos}(\Omega^l_{BR})$ , where $\Omega^l_{BR}$ is the physical AoD of the $l$-th path. Since $\Omega^l_{BR} \in [ -\pi,\pi  ]$, we have the full angle space range of $\left [ -1,1 \right ]$. In the top layer of the codebook, we equally divide the AoD full angle range into two codewords, so that the beam width of each codeword is one. Starting from a broad coverage pattern in the first layer, each subsequent layer doubles the number of codewords while halving the angular coverage per codeword, creating a hierarchical refinement from 1 to $S = \log_2 M$ layers \cite{qi2020hierarchical}.

Inspired by\cite{wu2019intelligent}, we deploy a two-stage approach to reduce the complexity of the joint MA-BS codebook and RIS phase shift optimization problem. This decoupled optimization strategy significantly reduces the training overhead compared to joint optimization methods, which is crucial for practical beam training systems facing limited coherence time and computational resources.

Specifically, the phase shifts at the RIS are optimized in the first stage by maximizing the sum of effective channel gains across users, solving the effective channel gain maximization problem using statistical CSI, applying MATLAB \texttt{fmincon} solver. The RIS phase optimization is performed once per coherence interval, avoiding the prohibitive computational cost of re-optimizing phases for each beam candidate during the hierarchical search. In the second stage, with the optimized RIS phase shift matrix $\boldsymbol\varphi$ fixed, we solve (P1) to obtain the proposed optimal blockage-aware codebook design for the MA-BS system. This sequential approach maintains the low-complexity advantage of hierarchical beam training while incorporating RIS optimization benefits.

\subsection{Statistical CSI for Stage-I Optimization}
Stage-I configures the RIS once per {statistical} (slow) timescale so as to precondition the cascaded BS--RIS--UE channel for the subsequent fast hierarchical beam training. This timescale separation is motivated by the fact that large-scale descriptors (angles, pathloss, spatial covariances, and blockage statistics) vary far more slowly than small-scale fading, and by the prohibitive cost of re-optimizing RIS phases every coherence interval. Concretely, the optimization is driven by second-order and large-scale information on the BS$ \to $RIS and RIS$ \to $UE links. Let $\mathbf R_{\text{BR}} = \mathbb E[\mathbf H_{\text{BR}} \mathbf H^H_{\text{BR}}]  \in \mathbb C^{N\times N}$ and $\mathbf R_{\text{RU},k} = \mathbb E[\mathbf H_{\text{RU}} \mathbf H^H_{\text{RU}}] \in \mathbb C^{N\times N}$ denote RIS-side spatial covariances for the BS$ \to $RIS and RIS$ \to $UE$_k$ channels, respectively, and let $\beta_{\text{BR}},\beta_{\text{RU},k}$ be their large-scale gains; in addition, sector-wise blockage probabilities $\pi_{\text{blk}}(\vartheta)$ summarize persistent angular occlusions and are later used only as masks during Stage-II pruning. These quantities enter the Stage-I objective through the expected effective gain (or a weighted sum across users). A convenient quadratic surrogate for this expected gain is 
\begin{align}
J_{\text{stat}} (\boldsymbol\varphi)\;&\triangleq\;\boldsymbol\varphi^H\mathbf Q\,\boldsymbol\varphi, \nonumber\\
\mathbf Q \;&\triangleq\;\sum_k w_k\Big(\beta_{\text{BR}}\beta_{\text{RU},k}\,\mathbf R_{\text{BR}}\odot \mathbf R_{\text{RU},k}^{\top}\Big),
\label{eq:Jstat}
\end{align}
where $\odot$ denotes the Hadamard product, $w_k \ge 0$ are service weights for user $k$, and the RIS phase shift for all $n$ is set as $|\varphi_n|=1$. RIS phases are then chosen as
\begin{align}
\widehat{\boldsymbol\varphi} &\in\arg\max_{|\varphi_n|=1}\;\boldsymbol\varphi^H\widehat{\mathbf Q}\,\boldsymbol\varphi, \nonumber\\
&\widehat{\mathbf Q}\text{ built from }\big(\widehat{\mathbf R}_{\text{BR}},\widehat{\mathbf R}_{\text{RU},k},\widehat{\beta},\widehat{\pi}_{\text{blk}}\big).
\label{eq:phi_hat_from_Qhat}
\end{align}
All statistical CSI is acquired over a {statistical window} $T_{\text{stat}}\!\gg$ coherence time using low-overhead RIS patterns and pilot beacons: the RIS cycles a small set of broad patterns while UEs (or the BS via reciprocity with calibration) return averaged received powers; from $S$ snapshots we form sample covariances $\widehat{\mathbf R}_{\text{BR}}$ and $\widehat{\mathbf R}_{\text{RU},k}$ via temporal averaging or subspace methods, and obtain large-scale gains $\widehat{\beta}$ by normalization; in parallel, an angular occupancy map $\widehat{\pi}_{\text{blk}}(\vartheta)$ is estimated by counting persistent energy drops per sector (and/or by fusing environment priors). Sampling error concentrates as $S$ grows; with i.i.d. snapshots and bounded fourth moments,
\begin{equation}
\|\widehat{\mathbf R}-\mathbf R\|_2=\mathcal{O}_{\mathbb P} \Big(\sigma^2\sqrt{\tfrac{N}{S}}\Big), 
\|\widehat{\mathbf R}-\mathbf R\|_F=\mathcal{O}_{\mathbb P} \Big(\sigma^2\tfrac{N}{\sqrt{S}}\Big),
\label{eq:cov_concentration}
\end{equation}
where we use $\mathbf R$ to uniformly represent the true covariance (either $\widehat{\mathbf R}_{\text{BR}}$ or $\widehat{\mathbf R}_{\text{RU},k}$) and $\widehat{\mathbf R}$ the corresponding sample estimate. Here $\mathcal{O}_{\mathbb P}(\cdot) $ denotes order in probability, indicating the estimation error scales at the given rate with high probability as $S$ increases. $\sigma^2$ reflects the snapshot SNR/dynamic range, and similar concentration holds for $\widehat{\beta}$ and for the sector frequencies underlying $\widehat{\pi}_{\text{blk}}$. To quantify how statistical-CSI error propagates to the Stage-I objective, let us define $\Delta\mathbf Q\!=\!\widehat{\mathbf Q}-\mathbf Q$. For any unit-modulus $\boldsymbol\varphi$,
\begin{equation}
\big|\boldsymbol\varphi^H \widehat{\mathbf Q}\,\boldsymbol\varphi - \boldsymbol\varphi^H \mathbf Q\,\boldsymbol\varphi\big|
\;\le\;\|\boldsymbol\varphi\|_2^2\,\|\Delta\mathbf Q\|_2
\;=\;N\,\|\Delta\mathbf Q\|_2,
\label{eq:quad_lipschitz}
\end{equation}
so if $\boldsymbol\varphi^\star\!\in\!\arg\max_{|\varphi_n|=1}\boldsymbol\varphi^H\mathbf Q\,\boldsymbol\varphi$ and $\widehat{\boldsymbol\varphi}$ solves \eqref{eq:phi_hat_from_Qhat}, then by Weyl’s inequality and \eqref{eq:quad_lipschitz}
\begin{align}
&\underbrace{\boldsymbol\varphi^{\star H}\mathbf Q\,\boldsymbol\varphi^\star}_{J_{\text{stat}}^\star}
- \underbrace{\widehat{\boldsymbol\varphi}^H\mathbf Q\,\widehat{\boldsymbol\varphi}}_{J_{\text{stat}}(\widehat{\boldsymbol\varphi})} \le \big(\lambda_{\max}(\mathbf Q)-\lambda_{\max}(\widehat{\mathbf Q})\big) \nonumber\\
&+ N\|\Delta\mathbf Q\|_2
\;\le\; (N{+}1)\,\|\Delta\mathbf Q\|_2,
\label{eq:Jstat_gap}
\end{align}
where $\lambda_{max}(\cdot)$ denotes the maximum eigenvalue, and first-order perturbation of \eqref{eq:Jstat} ties $\|\Delta\mathbf Q\|_2$ to the underlying covariance errors:
\begin{align}
\|\Delta\mathbf Q\|_2 &\;\le\; \sum_k |w_k|\Big(\beta_{\text{BR}}\beta_{\text{RU},k}\,\|\Delta\mathbf R_{\text{BR}}\|_2\,\|\mathbf R_{\text{RU},k}\|_2 \nonumber\\
&+\beta_{\text{BR}}\beta_{\text{RU},k}\,\|\mathbf R_{\text{BR}}\|_2\,\|\Delta\mathbf R_{\text{RU},k}\|_2\Big)\nonumber\\
&+\text{(small terms from }\widehat{\beta},\widehat{\pi}_{\text{blk}}).
\label{eq:Q_error_bound}
\end{align}
Combining \eqref{eq:cov_concentration}–\eqref{eq:Q_error_bound} yields the Stage-I value scaling
\begin{align}
J_{\text{stat}}^{\star}-J_{\text{stat}}(\widehat{\boldsymbol\varphi}) \;&\le\; (N{+}1)\,\mathcal{O}_{\mathbb P} \Big(\sqrt{\tfrac{N}{S}}\Big) \nonumber\\
&\text{(up to channel-dependent constants).}
\label{eq:Jstat_loss_scaling}
\end{align}
Finally, we connect the statistical objective to end performance. Let $\Gamma(\boldsymbol\varphi)$ denote the post-training SNR (or a deterministic equivalent) that is monotone in $J_{\text{stat}}(\boldsymbol\varphi)$. For a single stream $R(\boldsymbol\varphi)=\log_2(1+\Gamma(\boldsymbol\varphi))$, the mean rate loss due to using $\widehat{\boldsymbol\varphi}$ instead of $\boldsymbol\varphi^\star$ obeys
\begin{align}
\Delta R \triangleq R(\boldsymbol\varphi^\star)-R(\widehat{\boldsymbol\varphi})
\;&\le\;\frac{1}{\ln 2}\,\frac{\Delta\Gamma}{1+\Gamma(\boldsymbol\varphi^\star)}\nonumber\\
&\;\lesssim\;\frac{c}{\ln 2}\,\frac{(N{+}1)\,\|\Delta\mathbf Q\|_2}{1+\Gamma(\boldsymbol\varphi^\star)},
\label{eq:rate_gap_bound}
\end{align}
where $c$ absorbs the (positive) slope relating $\Gamma$ and $J_{\text{stat}}$. With total power $P_{\text{tot}}$ essentially unchanged by passive RIS phasing, the energy efficiency $\mathrm{EE}\!=\!R/P_{\text{tot}}$ satisfies
\begin{equation}
\Delta\mathrm{EE}=\frac{\Delta R}{P_{\text{tot}}}
\;\lesssim\;\frac{c}{P_{\text{tot}}\ln 2}\,\frac{(N{+}1)\,\|\Delta\mathbf Q\|_2}{1+\Gamma(\boldsymbol\varphi^\star)}.
\label{eq:ee_gap_bound}
\end{equation}
Equations \eqref{eq:rate_gap_bound}–\eqref{eq:ee_gap_bound} show that both rate and EE degradations are {linear} in $\|\Delta\mathbf Q\|_2$, which itself shrinks as $S^{-1/2}$; large nominal SNR further suppresses the impact through the $(1+\Gamma)^{-1}$ factor. Practically, $T_{\text{stat}}$ can span seconds to minutes in outdoor settings, making $S$ large and the bounds tight; Stage-II beam training is then carried out after fixing $\widehat{\boldsymbol\varphi}$ and selects the best instantaneous beams within pruned sectors, adding robustness to model mismatch; abrupt environmental changes (e.g., new blockage) can be handled by refreshing the blockage map and covariances with negligible added overhead via broad RIS patterns; and robust variants of \eqref{eq:phi_hat_from_Qhat} (e.g., max–min over $\|\mathbf Q-\widehat{\mathbf Q}\|_2\le \rho$) are available within the same computational envelope, trading a small nominal loss for guaranteed worst-case performance.

\subsection{Stage-II Blockage-Aware Hierarchical Codebook (HCB)}
\subsubsection{Blockage-Aware GS Codebook Design}

Let $\Phi_{\mathrm{ava}}$ denote the {available} (unblocked) angular set obtained from the blockage detection pipeline, and let $\Phi_{\mathrm{blk}}$ be its complement inside the field-of-view $\Phi_{\mathrm{FOV}}$: $\Phi_{\mathrm{ava}} \subseteq \Phi_{\mathrm{FOV}}$, and $\Phi_{\mathrm{blk}} \triangleq \Phi_{\mathrm{FOV}} \setminus \Phi_{\mathrm{ava}}$. For a uniform linear array with inter-element spacing $d=\lambda/2$ (extension to arbitrary arrays is straightforward), the array steering vector at spatial frequency $u=\sin\vartheta$ is
\begin{align}
    \mathbf a(u) = \frac{1}{\sqrt{M}}\big[1,\, e^{j\pi u},\, e^{j 2\pi u},\ldots, e^{j\pi(M-1) u}\big]^T.
\end{align}
Given a target sector $\Omega\subseteq\Phi_{\mathrm{ava}}$ (e.g., the coverage of a node at a given HCB layer), the conventional beam pattern-fitting objective seeks a weight vector $\mathbf w\in\mathbb C^M$ whose array factor matches a desired response $d(u)$ over $\Omega$ and suppresses leakage outside $\Omega$. Blockage awareness is enforced by {forbidding} or {penalizing} energy toward $\Phi_{\mathrm{blk}}$. Let $\mathcal U_{\Omega}=\{u_i\}_{i=1}^{N_\Omega}\subset\Omega$, $\mathcal U_{\mathrm{blk}}=\{v_j\}_{j=1}^{N_{\mathrm{blk}}}\subset\Phi_{\mathrm{blk}}$, and $\mathcal U_{\mathrm{sl}}=\{s_\ell\}_{\ell=1}^{N_{\mathrm{sl}}}\subset\Phi_{\mathrm{FOV}}\setminus(\Omega\cup\Phi_{\mathrm{blk}})$ be uniform or density-adaptive sampling grids for the main-lobe region, blocked directions, and sidelobe region, respectively. Define the steering dictionaries
\begin{align}
    \mathbf A_{\Omega} &\triangleq \big[\mathbf a(u_1),\ldots,\mathbf a(u_{N_\Omega})\big],\nonumber\\
\mathbf A_{\mathrm{blk}} &\triangleq \big[\mathbf a(v_1),\ldots,\mathbf a(v_{N_{\mathrm{blk}}})\big],\nonumber\\
\mathbf A_{\mathrm{sl}} &\triangleq \big[\mathbf a(s_1),\ldots,\mathbf a(s_{N_{\mathrm{sl}}})\big].
\end{align}
Let $\mathbf d\in\mathbb C^{N_\Omega}$ encode the desired response on $\mathcal U_{\Omega}$ (e.g., flat-top inside the sector), and let $P$ be the per-codeword power budget. In this case, we can derive a) \textbf{Hard-constrained blockage-aware design (exact nulling over $\Phi_{\mathrm{blk}}$)} to obtain,
\begin{equation}
\begin{aligned}
\min_{\mathbf w\in\mathbb C^M}\quad&
\underbrace{\left\|\mathbf A_{\Omega}^H\mathbf w - \mathbf d\right\|_2^2}_{\text{in-sector fit}}
+ \lambda_{\mathrm{sl}}\underbrace{\left\|\mathbf W_{\mathrm{sl}}\,\mathbf A_{\mathrm{sl}}^H\mathbf w\right\|_2^2}_{\text{sidelobe control}}\\[1mm]
\text{s.t.}\quad &
\underbrace{\mathbf A_{\mathrm{blk}}^H\mathbf w = \mathbf 0}_{\text{blockage avoidance on }\Phi_{\mathrm{blk}}},\qquad
\|\mathbf w\|_2^2 \le P .
\end{aligned}
\label{4ba}
\end{equation}
Here $\lambda_{\mathrm{sl}}\ge 0$ weights out-of-sector leakage and $\mathbf W_{\mathrm{sl}}\succcurlyeq 0$ (positive semidefinite) allows frequency-dependent side-lobe shaping. The constraint $\mathbf A_{\mathrm{blk}}^H\mathbf w=\mathbf 0$ forces {exact nulls} at all sampled blocked angles, thereby translating the geometric availability set directly into a linear constraint on $\mathbf w$. For b) \textbf{Soft-penalized blockage-aware design (leakage bounded or discouraged over $\Phi_{\mathrm{blk}}$)}, we use either an $\ell_2$ penalty or an $\ell_\infty$ cap:
\begin{equation}
\begin{aligned}
\min_{\mathbf w}\quad&
\left\|\mathbf A_{\Omega}^H\mathbf w - \mathbf d\right\|_2^2
+ \lambda_{\mathrm{blk}} \underbrace{\left\|\mathbf A_{\mathrm{blk}}^H\mathbf w\right\|_2^2}_{\text{blocked-angle leakage}} \nonumber\\
&+ \lambda_{\mathrm{sl}}\left\|\mathbf W_{\mathrm{sl}}\,\mathbf A_{\mathrm{sl}}^H\mathbf w\right\|_2^2\\
\text{s.t.}\quad & \|\mathbf w\|_2^2 \le P
\qquad\text{or}\qquad
\|\mathbf A_{\mathrm{blk}}^H\mathbf w\|_\infty \le \epsilon_{\mathrm{blk}} .
\end{aligned}
\label{4bb}
\end{equation}
Choosing $\lambda_{\mathrm{blk}}\!\uparrow\!\infty$ or $\epsilon_{\mathrm{blk}}\!\downarrow\!0$ recovers the hard-constraint behavior.

For the hard-constraint case, we can derive the closed-form formulation via projection. Let $\mathcal N\triangleq \mathrm{Null}(\mathbf A_{\mathrm{blk}}^H)$ and $\mathbf P_{\mathcal N}$ be the orthogonal projector onto $\mathcal N$: $\mathbf P_{\mathcal N} \;=\; \mathbf I - \mathbf A_{\mathrm{blk}}\big(\mathbf A_{\mathrm{blk}}^H\mathbf A_{\mathrm{blk}}\big)^{-1}\mathbf A_{\mathrm{blk}}^H$, if $\mathbf A_{\mathrm{blk}}$ is full column rank; else use pseudo-inverse. Feasible weights in \eqref{4ba} satisfy $\mathbf w=\mathbf P_{\mathcal N}\mathbf z$, where $\mathbf z \in \mathbb C^M$ is an auxiliary variable in the unconstrained space. Substituting and adding Tikhonov regularization $\mu\|\mathbf z\|_2^2$ to control power yields the unconstrained quadratic;
\begin{align}
    \min_{\mathbf z}\;\left\|\mathbf A_{\Omega}^H\mathbf P_{\mathcal N}\mathbf z - \mathbf d\right\|_2^2 
+ \lambda_{\mathrm{sl}}\left\|\mathbf W_{\mathrm{sl}}\,\mathbf A_{\mathrm{sl}}^H\mathbf P_{\mathcal N}\mathbf z\right\|_2^2
+ \mu\|\mathbf z\|_2^2 .
\label{eq:tik_powcontrl1}
\end{align}
Let $\widetilde{\mathbf A}_{\Omega} \triangleq\!\mathbf P_{\mathcal N}^H\mathbf A_{\Omega}$ and $\widetilde{\mathbf A}_{\mathrm{sl}}\!\triangleq\!\mathbf P_{\mathcal N}^H\mathbf A_{\mathrm{sl}}$. 

In that case, we rewrite the objective function$ f(\mathbf{z})$ in Equation (\ref{eq:tik_powcontrl1}),
\begin{equation}
    f(\mathbf{z}) = \left\|\widetilde{\mathbf A}_{\Omega}^H \mathbf z - \mathbf d\right\|_2^2 
+ \lambda_{\mathrm{sl}}\left\|\mathbf W_{\mathrm{sl}}\,\widetilde{\mathbf A}_{sl}^H\mathbf z\right\|_2^2
+ \mu\|\mathbf z\|_2^2 .
\label{eq:tik_powcontrl2}
\end{equation}
Then we firstly expand $\|\widetilde{\mathbf{A}}_{\Omega}^H \mathbf{z} - \mathbf{d}\|_2^2$  in (\ref{eq:tik_powcontrl2}):
\begin{align}
\|\widetilde{\mathbf{A}}_{\Omega}^H \mathbf{z} - \mathbf{d}\|_2^2 
&= (\widetilde{\mathbf{A}}_{\Omega}^H \mathbf{z} - \mathbf{d})^H(\widetilde{\mathbf{A}}_{\Omega}^H \mathbf{z} - \mathbf{d}) \nonumber\\
&= (\mathbf{z}^H \widetilde{\mathbf{A}}_{\Omega} - \mathbf{d}^H)(\widetilde{\mathbf{A}}_{\Omega}^H \mathbf{z} - \mathbf{d}) \nonumber\\
&= \mathbf{z}^H \widetilde{\mathbf{A}}_{\Omega} \widetilde{\mathbf{A}}_{\Omega}^H \mathbf{z} - \mathbf{z}^H \widetilde{\mathbf{A}}_{\Omega} \mathbf{d} - \mathbf{d}^H \widetilde{\mathbf{A}}_{\Omega}^H \mathbf{z} + \mathbf{d}^H \mathbf{d}.
\end{align}
Similarly, $\|\mathbf{W}_{\mathrm{sl}} \widetilde{\mathbf{A}}_{\mathrm{sl}}^H \mathbf{z}\|_2^2 = \mathbf{z}^H \widetilde{\mathbf{A}}_{\mathrm{sl}} \mathbf{W}_{\mathrm{sl}}^2 \widetilde{\mathbf{A}}_{\mathrm{sl}}^H \mathbf{z}$. Combining all terms, the objective becomes:
\begin{align}
f(\mathbf{z}) &= \mathbf{z}^H \widetilde{\mathbf{A}}_{\Omega} \widetilde{\mathbf{A}}_{\Omega}^H \mathbf{z} - \mathbf{z}^H \widetilde{\mathbf{A}}_{\Omega} \mathbf{d} - \mathbf{d}^H \widetilde{\mathbf{A}}_{\Omega}^H \mathbf{z} + \mathbf{d}^H \mathbf{d} \nonumber\\
&\quad + \lambda_{\mathrm{sl}} \mathbf{z}^H \widetilde{\mathbf{A}}_{\mathrm{sl}} \mathbf{W}_{\mathrm{sl}}^2 \widetilde{\mathbf{A}}_{\mathrm{sl}}^H \mathbf{z} + \mu \mathbf{z}^H \mathbf{z}.
\end{align}
The gradient of a quadratic form $\mathbf{z}^H \mathbf{Q} \mathbf{z} - \mathbf{z}^H \mathbf{b} - \mathbf{b}^H \mathbf{z}$ with respect to $\mathbf{z}$ is $2\mathbf{Q}\mathbf{z} - 2\mathbf{b}$. Setting the gradient to zero:
\begin{align}
\nabla_{\mathbf{z}} &= 2\left(\widetilde{\mathbf{A}}_{\Omega} \widetilde{\mathbf{A}}_{\Omega}^H + \lambda_{\mathrm{sl}} \widetilde{\mathbf{A}}_{\mathrm{sl}} \mathbf{W}_{\mathrm{sl}}^2 \widetilde{\mathbf{A}}_{\mathrm{sl}}^H + \mu \mathbf{I}\right)\mathbf{z} - 2\widetilde{\mathbf{A}}_{\Omega} \mathbf{d} = \mathbf{0}.
\label{eq:tik_grad}
\end{align}
Then, solving for $\mathbf z$ and recover $\mathbf w$ in (\ref{eq:tik_grad}), we can write,
\begin{align}
    \Big(\widetilde{\mathbf A}_{\Omega}\widetilde{\mathbf A}_{\Omega}^H
+ \lambda_{\mathrm{sl}}\,\widetilde{\mathbf A}_{\mathrm{sl}}\mathbf W_{\mathrm{sl}}^2\widetilde{\mathbf A}_{\mathrm{sl}}^H
+ \mu \mathbf I\Big)\mathbf z
 = \widetilde{\mathbf A}_{\Omega}\,\mathbf d; \mathbf w^\star = \mathbf P_{\mathcal N}\mathbf z.
\end{align}
Thus the hard-constraint solution equals the projection of a standard beam-pattern fit onto the null-space of the blocked-angle dictionary.

Next we investigate how GS can be implemented within the blockage-aware codebook design. Let $\mathbf B\triangleq\mathbf A_{\mathrm{blk}}$ and form a {blockage-aware basis} for $\mathcal N$ by GS ortho-normalization of the projected target dictionary: $\mathbf U \;=\; \mathrm{GS}\big(\mathbf P_{\mathcal N}\mathbf A_{\Omega}\big)\in\mathbb C^{M\times r},\qquad \mathbf U^H\mathbf U=\mathbf I$, where $\mathrm{GS}(\cdot)$ denotes the GS orthonormalization procedure, $r=\text{rank}(\mathbf P_{\mathcal N}\mathbf A_{\Omega}) \leq \min(M,N_\Omega)$ is the rank of $\mathbf P_{\mathcal N}\mathbf A_{\Omega}$ . We then restrict $\mathbf w$ to $\mathrm{span}(\mathbf U)$, i.e., $\mathbf w=\mathbf U\mathbf c$, and solve the reduced least-squares
\begin{equation}
    \min_{\mathbf c}\;\left\|\mathbf A_{\Omega}^H\mathbf U\,\mathbf c - \mathbf d\right\|_2^2
+ \lambda_{\mathrm{sl}}\left\|\mathbf W_{\mathrm{sl}}\,\mathbf A_{\mathrm{sl}}^H\mathbf U\,\mathbf c\right\|_2^2
+ \mu\|\mathbf c\|_2^2; \nonumber\\
\label{4bc}
\end{equation}
which is the GS-accelerated counterpart of \eqref{4ba}, reducing the problem from $M$ to $r$ dimensions. The optimal beamforming weights are recovered as $\mathbf{w}^\star = \mathbf{U}\mathbf{c}^\star$. Algorithmically, our GS iterations amount to:
\begin{enumerate}
  \item Build $\mathbf B=\mathbf A_{\mathrm{blk}}$ from $\Phi_{\mathrm{blk}}$ and compute $\mathbf P_{\mathcal N}$ (or implicitly maintain orthogonality via GS).
  \item Project the target dictionary: $\widehat{\mathbf A}_{\Omega}=\mathbf P_{\mathcal N}\mathbf A_{\Omega}$, then obtain an orthonormal basis $\mathbf U=\mathrm{GS}(\widehat{\mathbf A}_{\Omega})$.
  \item Solve \eqref{4bc} for $\mathbf c^\star$ (small $r\times r$ system), set $\mathbf w^\star=\mathbf U\mathbf c^\star$, and (if needed) normalize or map to constant-modulus phases by $\angle(\mathbf w^\star)$.
\end{enumerate}
The hard-constraint formulation offers clear blockage guarantees via exact nulling over $\Phi_{\mathrm{blk}}$, whereas the soft formulation permits a small, tunable leakage (through $\lambda_{\mathrm{blk}}$ or $\epsilon_{\mathrm{blk}}$) in exchange for additional degrees of freedom to shape the main lobe. The GS step is not heuristic: it explicitly constructs an orthonormal basis for the blocked-angle null-space and then solves a reduced, well-conditioned least-squares problem within that subspace before mapping the solution back to the full array space. Finally, the design weights and thresholds $(\lambda_{\mathrm{blk}},\lambda_{\mathrm{sl}},\epsilon_{\mathrm{blk}},\mu)$ can be tied to peak/average beam-pattern specifications and selected once per stage via lightweight validation sweeps or closed-form approximations so as to maintain a consistent target side lobe level across layers.

\subsubsection{GS Adaptation: Blockage-Aware Target Pattern and Iterative Steps}
Let the field-of-view be $\Phi_{\mathrm{FOV}}$ and define the {blocked} and {available} angle sets
\begin{align}
    \Phi_{\mathrm{blk}} \;\triangleq\; \bigcup_{o}\big[\phi^{\min}_{r,m,o},\,\phi^{\max}_{r,m,o}\big]; \Phi_{\mathrm{ava}} \;\triangleq\; \Phi_{\mathrm{FOV}}\setminus \Phi_{\mathrm{blk}} .
\end{align}
Consider a node in the hierarchical codebook indexed by $(s,\ell)$, where $s$ denotes the hierarchy level and $\ell$ the beam index, with intended coverage sector $\Omega_{s,\ell}\subseteq\Phi_{\mathrm{ava}}$, the array factor of a weight vector $\mathbf w\in\mathbb C^M$ is $F_{\mathbf w}(u)=\mathbf a(u)^H\mathbf w$ and the original (16) target power pattern is $g_{\mathbf w}(u)=|F_{\mathbf w}(u)|^2$ on $\Phi_{\mathrm{FOV}}$. We {modify} this target to enforce blockage awareness by prescribing a {piecewise} desired magnitude profile:
\begin{equation}
\;g_{\mathbf w}^{\mathrm{aware}}(u)\;=\;
\begin{cases}
G_{\mathrm{in}}(u), & u\in \Omega_{s,\ell}\subseteq\Phi_{\mathrm{ava}},\\[2pt]
0,                  & u\in \Phi_{\mathrm{blk}},\\[2pt]
G_{\mathrm{sl}}(u), & u\in \Phi_{\mathrm{FOV}}\setminus\big(\Omega_{s,\ell}\cup\Phi_{\mathrm{blk}}\big),
\end{cases}
\label{4be}
\end{equation}
with $G_{\mathrm{in}}(u)\ge 0,\; G_{\mathrm{sl}}(u)\in[0,\gamma_{\mathrm{sl}}]$. Hence, inside the intended sector we seek a flat-top or designed gain pattern $G_{\mathrm{in}}$, in blocked directions we {force zero} desired gain, and elsewhere we cap leakage via $G_{\mathrm{sl}}$ (e.g., a small constant $\gamma_{\mathrm{sl}}$ or a taper). Discretizing $\Phi_{\mathrm{FOV}}$ on a grid $\mathcal U=\{u_i\}_{i=1}^{N_u}$ with steering matrix $\mathbf A=[\mathbf a(u_1),\ldots,\mathbf a(u_{N_u})]$, we collect the target magnitudes as
\begin{align}
\mathbf g^{\mathrm{aware}}\;&\triangleq\;\big[g_{\mathbf w}^{\mathrm{aware}}(u_1),\ldots,g_{\mathbf w}^{\mathrm{aware}}(u_{N_u})\big]^T, \nonumber\\
\boldsymbol{\rho}\;&\triangleq\;\sqrt{\mathbf g^{\mathrm{aware}}}\;\;(\text{elementwise}),\nonumber\\
&\mathcal I_{\mathrm{in}},\mathcal I_{\mathrm{blk}},\mathcal I_{\mathrm{sl}}\subset\{1,\ldots,N_u\}.
\label{4bd}
\end{align}
where $\mathcal I_{\mathrm{in}},\mathcal I_{\mathrm{blk}},\mathcal I_{\mathrm{sl}} \subset\{1,\ldots,N_u\}$ correspond to samples in the target sector $\Omega_{s,\ell}$, blockage set $\Phi_{\mathrm{blk}}$, and remaining sidelobe region $\Phi_{\mathrm{FOV}}\setminus\big(\Omega_{s,\ell}\cup\Phi_{\mathrm{blk}}\big)$, respectively. Starting from any $\mathbf w^{(0)}$ (e.g., random phases), for $t=0,1,\ldots,I_{\max}-1$ iterate:
\begin{enumerate}
\item {Forward map (array domain $\to$ angular domain):}
$
\mathbf F^{(t)} \;=\; \mathbf A^{H}\mathbf w^{(t)} \in \mathbb C^{N_u},\;
F^{(t)}_i = \mathbf a(u_i)^H\mathbf w^{(t)}.
$
\item {Blockage-aware amplitude enforcement:}
replace the magnitude of $\mathbf F^{(t)}$ by the prescribed targets while {retaining phase}:
\begin{equation}
\widetilde F^{(t)}_i \;=\;
\begin{cases}
\sqrt{G_{\mathrm{in}}(u_i)}\,e^{j\angle F^{(t)}_i}, & i\in \mathcal I_{\mathrm{in}}, \\[2pt]
0,                                                  & i\in \mathcal I_{\mathrm{blk}}, \\[2pt]
\min\!\big\{|F^{(t)}_i|,\,\sqrt{G_{\mathrm{sl}}(u_i)}\big\}\,e^{j\angle F^{(t)}_i}, & i\in \mathcal I_{\mathrm{sl}} .
\end{cases}
\label{4bf}
\end{equation}
{Hard exclusion alternative:} instead of zeroing blocked samples, simply {remove} $i\in\mathcal I_{\mathrm{blk}}$ from the constraint set; this is equivalent to enforcing \eqref{4bf} with entries omitted and corresponds to domain exclusion.
\item {Inverse map (least-squares back to array weights):} $\mathbf w^{(t+\tfrac12)} \;=\; \arg\min_{\mathbf w}\,\big\|\mathbf A^{H}\mathbf w - \widetilde{\mathbf F}^{(t)}\big\|_2^2 \Rightarrow \mathbf w^{(t+\tfrac12)} \;=\; \big(\mathbf A\mathbf A^{H}+\mu\mathbf I\big)^{-1}\mathbf A\,\widetilde{\mathbf F}^{(t)}$, with a small $\mu\!\ge\!0$ (Tikhonov) to improve conditioning if needed.
\item {Element-wise amplitude constraint (constant-modulus or power):}
\begin{align}
\mathbf w^{(t+1)} \;=\; 
\begin{cases}
&\sqrt{\tfrac{P}{M}}\; e^{j\,\angle(\mathbf w^{(t+\tfrac12)})},\\ 
& \text{(constant-modulus array)},\\
&\sqrt{\tfrac{P}{\|\mathbf w^{(t+\tfrac12)}\|_2^2}}\;\mathbf w^{(t+\tfrac12)},\\ 
& \text{(total-power constraint)}.
\end{cases}
\label{eq:cm_projection}
\end{align}
\end{enumerate}
In summary, Steps~\eqref{4be}–\eqref{eq:cm_projection} implement a classical GS scheme of amplitude replacement in the Fourier (angular) plane followed by projection in the spatial plane, with the key difference that the angular-plane amplitudes are {explicitly masked} by the blockage-aware target $g_{\mathbf w}^{\mathrm{aware}}$ from~\eqref{4bf}: blocked angles either receive a {zero} desired magnitude (hard blocking) or are {excluded} from the enforcement set (domain exclusion), both choices being fully determined by the indicator sets $(\mathcal I_{\mathrm{in}},\mathcal I_{\mathrm{blk}},\mathcal I_{\mathrm{sl}})$ derived from (15); specializing per node $(s,\ell)$ of the hierarchy, we use $\Omega_{s,\ell}$ in~\eqref{4bf} with the same $\Phi_{\mathrm{blk}}$, so the iteration yields $\mathbf w_{s,\ell}$ that directs no energy into $\Phi_{\mathrm{blk}}$ (hard case) or guarantees bounded side lobes there (soft case), while the element-wise amplitude constraint is enforced by~\eqref{eq:cm_projection}, thereby precisely defining the modified target pattern $g_{\mathbf w}^{\mathrm{aware}}$ and its role within each GS iteration.

\section{Performance Analysis}

This section consolidates the algorithmic developments of Sec.\ref{sec:algo} into quantitative performance guarantees and costs. We first translate the proposed two-stage pipeline into computational terms by deriving closed-form expression at each hierarchy layer and across the entire hierarchical structure complexity for the blockage-aware GS codebook and by amortizing all design-time operations (RIS optimization, blockage-map updates, and hierarchical codebook construction) over the environment’s coherence, yielding a per-interval (or per-slot) total complexity that cleanly separates {run-time sensing}---proportional to the number of beam evaluations $B_{\text{method}}$, users $K$, and pilot length $L_p$---from {infrequent design} costs. We then analyze the convergence of the adapted GS routine viewed as alternating projections between an affine least-squares (LS) set and a nonconvex amplitude-constraint set, establish a monotone decrease of a masked beam-pattern-fit surrogate, and justify the iteration cap $I_{\max}$ via stopping criteria that stabilize in a few tens of iterations across hierarchy layers. Throughout, we benchmark against DFT, Subspace Codebook (SC), and traditional hierarchical training, use the timing and power parameters in Table~1 to ground the temporal accounting with the slot duration $T$ and mechanical move time $\tau$, and report all results per coherence interval while explicitly exposing refresh factors for (i) statistical CSI/RIS phases, (ii) blockage map $\Phi_{\mathrm{ava}}$, and (iii) GS codebook rebuilds. This organization links computational effort to achievable rate and energy efficiency, preparing the ground for the numerical validation.

\subsection{Complexity Analysis}

We decompose the total complexity into three components (i) Stage - I: RIS phase-shift Optimization, (ii) Stage - II: Hierarchical GS codebook across all layers and (iii) Run-time Beam Training (measurement followed by decoding and decision).

\subsubsection{Stage-I: RIS Phase-Shift Optimization}
Stage-I maximizes a smooth objective in the RIS phases $\Phi=\mathrm{diag}(\boldsymbol\varphi)$ using statistical CSI via MATLAB \texttt{fmincon}, and is executed once per coherence interval before beam training. The system model uses
\begin{align}\label{4aa}
    \mathbf{H}_{k} = \mathbf{H}_{\text{RU},k} ^{H}\,\Phi\,\mathbf{H}_{\text{BR}}, \qquad \mathbf{H}_{\text{BR}}\in\mathbb{C}^{N\times M},\; \mathbf{H}_{\text{RU},k} \in\mathbb{C}^{N\times 1}.
\end{align}
\texttt{fmincon} solves a sequence of Karush–Kuhn–Tucker (KKT) linear systems of size approximately $(N{+}m)$, where $m$ is the number of active (nonlinear/linear) constraints (negligible when we have only box constraints for phases). With $I_{\text{RIS}}$ iterations, the per-iteration cost for full (dense) matrix operations is dominated by the KKT solve (cubic) plus one objective/derivative evaluation: $\mathcal{C}_{\text{RIS}} = \mathcal{O} \big(I_{\text{RIS}}\big((N{+}m)^3 + C_{\text{model}}\big)\big)$. Using $h_k$, one model evaluation applies a diagonal and a dense product across users/terms, which is linear in the model size: $C_{\text{model}}=\Theta(MN)$, where $\Theta(\cdot)$ is an arbitrary linear function. Two practically relevant solver variants are: a) \textbf{Analytic (or auto-diff) gradients / quasi-Newton Hessian:}
\begin{align}
    \;\mathcal{C}_{\text{RIS}}=\mathcal{O} \big(I_{\text{RIS}}(N{+}m)^3 + I_{\text{RIS}}MN\big)\;
    \label{4ab}
\end{align}
and b) \textbf{Finite-difference gradients (no analytic gradient):}
\begin{align}
\;\mathcal{C}_{\text{RIS}}=\mathcal{O} \big(I_{\text{RIS}}(N{+}m)^3 + I_{\text{RIS}}\,M N^2\big)\;
    \label{4ac}
\end{align}
In both cases, the {cubic KKT solve} dominates for sufficiently large $N$, yielding the simplified bound used later:
\begin{align}
    {\;\mathcal{C}_{\text{RIS}}=\mathcal{O} \big(I_{\text{RIS}}N^3\big)\;}
    \label{4ad}
\end{align}

\subsubsection{Stage-II: Hierarchical GS Codebook (HCB)}
From the GS procedure, generating a {single} codeword costs
\begin{align}
   \;\mathcal{C}_{\text{GS,1}}=\mathcal{O} \big(I_{\max}M^{2}+M^{3}\big)\;, 
    \label{4ae}
\end{align}
where $M^3$ accounts for the least-squares/normal-equation solve and $I_{\max}M^2$ for per-iteration operations. The HCB is a {binary tree} with $S=\log_2 M$ layers, having $2^{s-1}$ codewords at layer $s$. Therefore the {total} GS cost across {all} layers/codewords is
\begin{align}
    \mathcal{C}_{\text{HCB}}
    &= \sum_{s=1}^{S} 2^{\,s-1}\,\mathcal{C}_{\text{GS,1}}
     = (2^{S}-1)\,\mathcal{C}_{\text{GS,1}}
     = (M-1)\,\mathcal{C}_{\text{GS,1}} \nonumber\\ 
     &(\text{since } S=\log_2 M), \\
    \;\mathcal{C}_{\text{HCB}}
    &= \mathcal{O} \big((M-1)(I_{\max}M^2+M^3)\big)\nonumber\\
     &=\mathcal{O} \big(I_{\max}M^3+M^4\big)\;
    \label{4af}
\end{align}
The blockage detection uses geometric tests and angular unions ((12)–(15)). Given $O$ blockages and $N$ RIS elements, evaluating the geometric predicates for all triples $(m,n,o)$ costs $\Theta(MNO)$. Merging the $O$ intervals per MA via sort-and-sweep is $\Theta(O\log O)$ per MA, giving $\mathcal{C}_{\text{blk}}=\mathcal{O} \big(MNO + M\,O\log O\big)$. Beam-coverage rotation \ref{eq2} requires $\Theta(M)$ per node. Across the whole hierarchy,
\begin{align}
{\;\mathcal{C}_{\text{rot,all}}=\Theta \big(M(2^{S}-1)\big)=\Theta(M^2)\;}
    \label{4ah}
\end{align}
which is negligible compared with \eqref{4af}.

\subsubsection{Run-Time Beam Training}

During one beam evaluation, the BS transmits a pilot with beamforming vector $\mathbf{w}$; at UE $k$,
\begin{align}
    y_k = \sqrt{p}\,\mathbf{H}_{k}^H \mathbf{w} + n_k 
    \Rightarrow
    \gamma_k = \frac{p\lvert  \mathbf{H}^H_{k} \mathbf{w}\rvert^2}{\sigma^2} 
\end{align}
(energy per SNR numerical value) followed by either a hierarchical approach using local pairwise comparisons between nodes at the same tree level, or an exhaustive approach that evaluates all candidates globally. The {computation} per beam evaluation is dominated by the length-$L_p$ accumulation plus a constant-time magnitude/compare: $\Theta(L_p)$ per UE. Hence,
\begin{equation}
    {\;\mathcal{C}_{\text{eval}}=\Theta(KL_p)\;}, 
    {\;\mathcal{C}_{\text{train}}(\text{method})=\Theta \big(B_{\text{method}}\,K\,L_p\big)\;}
    \label{4ai}
\end{equation}
where $B_{\text{method}}$ is the number of beam evaluations within one coherence interval. The mechanical move time $\tau$ affects duration but not flops. For our proposed blockage-aware hierarchical technique, pruning removes blocked branches and therefore, can be deduced as,
    \begin{align}
        {\;S \ \le\ B_{\text{Prop}} \ \le\ 2S\;} 
        \label{4aj}
    \end{align}
where $S=\log_2 M$, \textbf{$ B_{\text{Prop}}$} refers to the number of beam evaluations for our proposed method.

\subsubsection{Total Complexity} 
Let $R$ be the average number of coherence intervals between HCB regenerations and $R_{\text{RIS}}$ for RIS updates. Combining run-time training with design-time terms, we obtain :
\begin{equation}
\begin{aligned}
\mathcal{C}_{\text{total/interval}}&(\text{method})
=\underbrace{\Theta\!\big(B_{\text{method}}\,K\,L_p\big)}_{\text{beam training (run-time)}} \\
&+\underbrace{\dfrac{\mathcal{O}(I_{\max}M^3+M^4)}{R}}_{\text{GS/HCB over all layers}} 
+\underbrace{\dfrac{\mathcal{O}(MNO + M\,O\log O)}{R}}_{\text{blockage detection}} \\
&+\underbrace{\dfrac{\mathcal{O}(I_{\text{RIS}}N^3)}{R_{\text{RIS}}}}_{\text{Stage-I RIS}} 
+\underbrace{\Theta(M^2)}_{\text{beam-rotation bookkeeping}}
\end{aligned}
\label{4ak}
\end{equation}
Using $B_{\text{Prop}}\le 2\log_2 M$ in the beam training term, we can summarize \eqref{4ak} for our proposed technique as,
\begin{align}
\mathcal{C}_{\text{total/interval}} 
&= \Theta \big((\log_2 M)\,K\,L_p\big) \nonumber\\
&+ \frac{\mathcal{O}(I_{\max}M^3+M^4)}{R} 
+ \frac{\mathcal{O}(MNO+M\,O\log O)}{R} \nonumber\\
&+ \frac{\mathcal{O}(I_{\text{RIS}}N^3)}{R_{\text{RIS}}}
+ \Theta (M^2).
\label{4al}
\end{align}
Physically, \eqref{4al} states that the per-interval compute splits into a real-time sensing term $\Theta((\log_2 M)K L_p)$ that grows only logarithmically with array size $M$ and linearly with users, plus design terms $\frac{\mathcal{O}(I_{\max}M^3+M^4)}{R}$ (beam synthesis), $\frac{\mathcal{O}(MNO+M\,O\log O)}{R}$ (blockage geometry), and $\frac{\mathcal{O}(I_{\text{RIS}}N^3)}{R_{\text{RIS}}}$ (RIS configuration). When $R,R_{\text{RIS}}\!\gg\!1$ the sensing term dominates (training-limited regime), whereas very large $N$ or frequent updates shift dominance to the cubic RIS KKT solver and — if $R$ is small or $M$ is large—to the GS $M^4$ hierarchy build.

\subsection{Convergence and Update-Rate Analysis}
\label{conv}
The iterative procedure used in Stage~II alternates between a LS update in the spatial domain and amplitude enforcement in the angular domain, with masking induced by the blockage-aware target pattern. Although the beam-pattern fitting objective in isolation is an $\ell_2$-norm and therefore convex, the {overall} problem solved by the adapted GS routine is {nonconvex}: after the LS step, we enforce magnitudes in the angular plane according to the masked target $g_{\mathbf w}^{\mathrm{aware}}$(maintaining phases or excluding blocked samples) and then project the array weights onto a constant–modulus (or fixed-power) set. Consequently, the algorithm is best viewed as an alternating-projections / projected-LS method between an affine set $\mathcal S=\{\mathbf w:\,\mathbf A^H\mathbf w\approx \widetilde{\mathbf F}\}$ and a nonconvex constraint set $\mathcal C$ (masked angular magnitudes plus element-wise amplitude constraints).

To formalize the convergence behavior, let $\mathcal I_{\mathrm{in}},\mathcal I_{\mathrm{blk}},\mathcal I_{\mathrm{sl}}$ be the angular sample indices for the in-sector, blocked, and sidelobe regions, respectively, and let $\mathbf g^{\mathrm{aware}}$ collect the piecewise target magnitudes with $\boldsymbol\rho=\sqrt{\mathbf g^{\mathrm{aware}}}$ (elementwise). At iteration $t$, define the masked fit error
\[
E_t \;\triangleq\; \big\|\mathbf A^H\mathbf w^{(t)} - \widetilde{\mathbf F}^{(t)}\big\|_2^2,
\]
where $\widetilde{\mathbf F}^{(t)}$ is obtained by replacing the magnitude of $\mathbf A^H\mathbf w^{(t)}$ by the prescribed targets on $\mathcal I_{\mathrm{in}}\cup\mathcal I_{\mathrm{sl}}$ and by zero (or domain exclusion) on $\mathcal I_{\mathrm{blk}}$. The LS step computes $\mathbf w^{(t+\tfrac12)}$ with a small Tikhonov parameter $\mu \ge 0$ to improve conditioning if needed. By optimality of $\mathbf w^{(t+\tfrac12)}$, the surrogate energy is non-increasing across the half-step, i.e., $E_{t+\tfrac12}\le E_t$. The subsequent projection onto the constant–modulus (or normalized–power) set, given by (\ref{eq:cm_projection}) is non-expansive in the Euclidean norm, so the composite iteration renders $\{E_t\}$ bounded below by $0$ and monotone nonincreasing, hence convergent. Any limit point $\mathbf w^\odot$ of the iterates satisfies a GS-type fixed-point condition: its angular spectrum matches the masked magnitudes on $\mathcal I_{\mathrm{in}}\cup\mathcal I_{\mathrm{sl}}$ and is zero (or unconstrained, under domain exclusion) on $\mathcal I_{\mathrm{blk}}$, while $\mathbf w^\odot$ lies on the amplitude-constraint set in the array domain. Thus, while the problem is nonconvex, the objective surrogate used to monitor progress is provably decreasing and the algorithm converges to a stationary point consistent with the enforced masks and hardware constraints.

For reproducibility and to justify the iteration budget, we adopt two stopping tests per node. First, we require a small relative decrease of the masked fit,
\begin{align}
    \frac{\big\|\mathbf A^H\mathbf w^{(t)}-\widetilde{\mathbf F}^{(t)}\big\|_2^2-\big\|\mathbf A^H\mathbf w^{(t+1)}-\widetilde{\mathbf F}^{(t+1)}\big\|_2^2}{\big\|\mathbf A^H\mathbf w^{(t)}-\widetilde{\mathbf F}^{(t)}\big\|_2^2}
\;<\;\varepsilon,
\end{align}
for $\varepsilon\in[10^{-4},10^{-3}]$ and, second, we target a normalized beampattern mean squared error below a level induced by the sidelobe cap $\gamma_{\mathrm{sl}}$ on the active grid $\mathcal I_{\mathrm{in}}\cup\mathcal I_{\mathrm{sl}}$. In all scenarios considered, both criteria stabilize within a few tens of iterations across layers; we therefore set a conservative iteration cap $I_{\max}$ that exceeds the largest observed count and use it in the complexity statements $O(I_{\max}M^2+M^3)$ for per-codeword generation and $O(I_{\max}M^3+M^4)$ for full-hierarchy construction. Empirically, blockage-aware pruning accelerates stabilization: by eliminating infeasible directions early, the enforced angular domain becomes smaller and more localized as we descend the hierarchy, which improves conditioning of $\mathbf A\mathbf A^H$ on the active grid and speeds up the LS–projection loop. In the hard-nulling variant, if the blocked dictionary $\mathbf A_{\mathrm{blk}}$ becomes high-rank relative to $M$ and the feasible subspace $\mathrm{Null}(\mathbf A_{\mathrm{blk}}^H)$ narrows, the projector/Tikhonov formulation mitigates any slow-down by keeping iterates close to the feasible region and improving numerical conditioning. Furthermore, warm-starting each child node with its parent's solution accelerates convergence. 

To visually support the analysis, we include a convergence plot seen as Fig. \ref{cov1} (in the supplementary) that traces the masked fit $\|\mathbf A^H\mathbf w^{(t)}-\widetilde{\mathbf F}^{(t)}\|_2^2$ versus iteration index $t$ across several blockage densities and node depths, showing the characteristic rapid drop followed by a plateau; the chosen $I_{\max}$ is placed beyond the knee of these curves in all cases.

Finally, we relate the iterative design overhead to environmental coherence by separating three refresh timescales per training slot of duration $T$. (i) {Statistical CSI / RIS phases} (Stage~I) are refreshed every $R_{\mathrm{RIS}}$ slots, yielding an amortized per-slot complexity of $\mathcal{O}(I_{\mathrm{RIS}}N^3)/R_{\mathrm{RIS}}$. (ii) The {blockage map} $\Phi_{\mathrm{ava}}$ (derived from the geometric tests) is refreshed every $R_{\mathrm{blk}}$ slots, contributing $\mathcal{O}(MNO+M\,O\log O)/R_{\mathrm{blk}}$ per slot. (iii) The {GS-based codebook} (Stage~II) is rebuilt every $R_{\mathrm{cb}}$ slots—or event-triggered when the Jaccard change (measuring angular overlap change) of $\Phi_{\mathrm{ava}}$ exceeds a threshold—with amortized cost $\mathcal{O}(I_{\max}M^3+M^4)/R_{\mathrm{cb}}$. Adding the run-time training term $\Theta(B_{\text{method}}KL_p)$, the per-slot total matches the system-level accounting given earlier: training dominates when $R_{\mathrm{RIS}},R_{\mathrm{blk}},R_{\mathrm{cb}}\gg1$ (typical, since large-scale covariances and blockage statistics vary on a seconds–minutes scale), whereas faster dynamics call for increasing $R_{\mathrm{blk}}^{-1}$ while keeping $R_{\mathrm{cb}}^{-1}$ modest by reusing the existing hierarchy and updating only affected branches. With the quantitative parameters used in our evaluations (e.g., slot duration $T$ and mechanical move time $\tau$), the amortized design costs remain well below the run-time sensing load for moderate refresh rates, so iterative codebook updates do not negate the beam-training savings delivered by the blockage-aware hierarchy.

\subsection{Two-stage Optimization Performance}

\begin{figure}[!t]
\centering
        \includegraphics[width=1.01\linewidth]{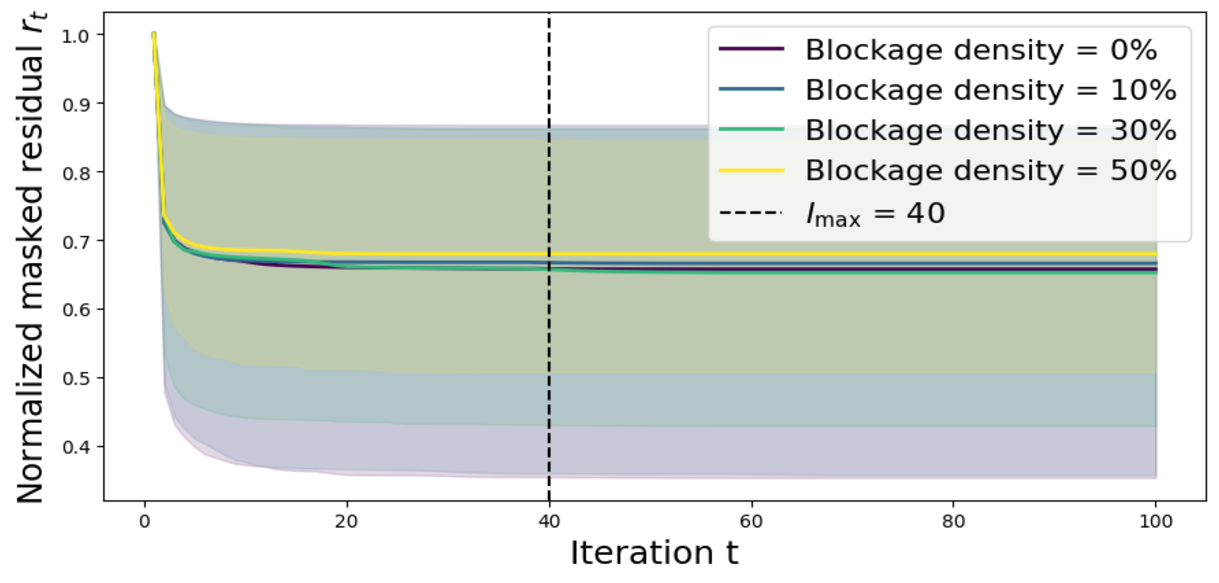}
        \caption{Empirical convergence of the blockage-aware GS algorithm under different blockage densities. The figure shows the normalized masked residual $r_t = E_t / E_0$ versus iteration index $t$ for blockage densities of $\{0\%, 10\%, 30\%, 50\%\}$. Solid lines represent mean trajectories, while shaded regions depict the Interquartile Range (IQR) across multiple random blockage pattern realizations for each density level. The vertical dashed line marks the conservative iteration cap $I_{\max}{=}40$. All curves exhibit the characteristic GS convergence profile—rapid initial decrease followed by a plateau—indicating stable convergence across scenarios.}
        \vspace{-4mm}
\label{cov1}
\end{figure}

\begin{figure}[t]
\centering
        \includegraphics[width=1.01\linewidth]{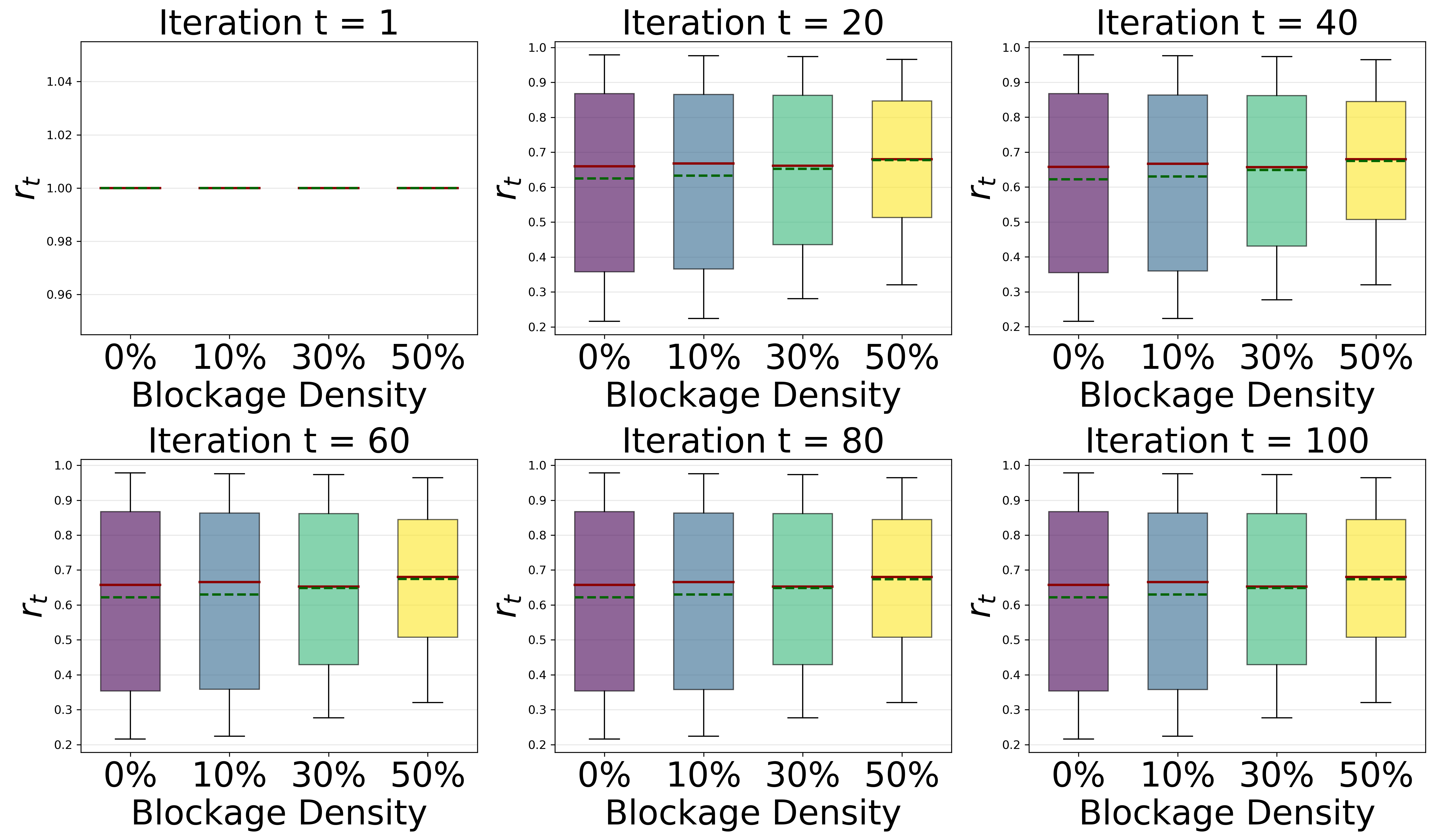}
        \caption{Distributional analysis of normalized residuals $r_t$ at selected iterations for blockage densities of 0\%, 10\%, 30\%, and 50\%. Each box plot synthesizes 240 independent sector realizations, with solid red lines marking medians, dashed green lines indicating means, and box edges spanning the IQR. Rapid convergence is evident: median residuals decrease from 1 to approximately 0.66 by $t=20$ and remain stable thereafter (variation $< 0.01$ through $t=100$), indicating that the GS algorithm reaches a stable operating region within 20 iterations. }
        \vspace{-4mm}
\label{cov2}
\end{figure}

We proposed two mathematically distinct ways to incorporate blockage into Stage-II: a non-iterative projector–LS path and an iterative GS path. The former enforces hard nulls by projecting the target dictionary $\mathbf A_{\Omega}$ onto the blocked-angle null-space $\mathcal N=\mathrm{Null}(\mathbf A_{\mathrm{blk}}^H)$ via $\mathbf P_{\mathcal N}$ and then solving a (regularized) least-squares problem for $\mathbf z$ (or coefficients $\mathbf c$ in an orthonormal basis $\mathbf U=\mathrm{GS}(\mathbf P_{\mathcal N}\mathbf A_{\Omega})$), which yields in one shot a convex LS solution within $\mathcal N$ but {does not} satisfy the analog constant-modulus (CM) hardware constraint without a further nonconvex projection. By contrast, the iterative GS path alternates between enforcing a blockage-aware magnitude mask in the angular domain and enforcing the CM (or fixed-power) constraint in the array domain; to reconcile blockage nulls with CM in our final design, we explicitly insert the null-space projector into each iteration so that the codebook generation we {actually use} is the following composed GS loop: (i) angular amplitude enforcement produces $\widetilde{\mathbf F}^{(t)}$ by setting the magnitudes on in-sector samples to the desired profile, zeroing blocked samples, and capping side-lobes; (ii) array LS back-projection computes $\widehat{\mathbf w}^{(t+\frac12)}=(\mathbf A\mathbf A^H+\mu\mathbf I)^{-1}\mathbf A\,\widetilde{\mathbf F}^{(t)}$; (iii) blockage-null enforcement projects to the feasible subspace $\mathbf w^{(t+\frac12)}=\mathbf P_{\mathcal N}\widehat{\mathbf w}^{(t+\frac12)}$ with $\mathbf P_{\mathcal N}=\mathbf I-\mathbf A_{\mathrm{blk}}(\mathbf A_{\mathrm{blk}}^H\mathbf A_{\mathrm{blk}})^{-1}\mathbf A_{\mathrm{blk}}^H$; and (iv) array amplitude constraint applies the CM projection $\mathbf w^{(t+1)}=\sqrt{P/M}\,e^{j\angle(\mathbf w^{(t+\frac12)})}$ (or power normalization). We pre-factor $(\mathbf A\mathbf A^H+\mu\mathbf I)$ and compute a reduced-size QR/SVD of $\mathbf A_{\mathrm{blk}}$ once per node so that applying Step (ii) and the projector $\mathbf P_{\mathcal N}=\mathbf I-\mathbf Q_{\mathrm{blk}}\mathbf Q_{\mathrm{blk}}^H$ in Step (iii) is $\Theta(M^2)$ per iteration after an initial $\Theta(M^3)$ setup; consequently, the per-codeword complexity is $O(M^3)+O(I_{\max}M^2)$ and the total for the full hierarchy is $O(I_{\max}M^3+M^4)$, which is the complexity reported in Sec.~IV and is {specific} to the iterative path. The one-shot projector–LS path is kept in our analysis only as an auxiliary tool: it serves as a warm start for the iterative loop (set $\mathbf w^{(0)}$ to the LS solution before CM projection) and as a digital/relaxed-amplitude baseline when CM is not required, in which case the cost reduces to $O(M^3)$ per codeword (factorizations) and $O(M^4)$ for the hierarchy. With this choice, the convergence discussion in Sec. \ref{conv} applies directly to the composed GS iteration above: the masked beampattern-fit surrogate $\|\mathbf A^H\mathbf w^{(t)}-\widetilde{\mathbf F}^{(t)}\|_2^2$ is monotonically non-increasing through the LS step, the subsequent projections (null-space and CM) are non-expansive, and the iterates converge to a GS-type stationary point that simultaneously satisfies the blockage mask, hard nulls, and the CM array constraint; the empirical curves (Fig.~\ref{cov1} and Fig.~\ref{cov2}) justify a single conservative $I_{\max}$ that upper-bounds the observed iteration counts across layers and blockage densities, and they also confirm that pruning accelerates stabilization.

Fig. \ref{cov1} depicts the normalized masked residual $r_t=E_t/E_0$ versus iteration index $t$ for blockage densities $\{0\%,10\%,30\%,50\%\}$, where each curve aggregates results over multiple random blockage patterns. The shaded bands capture the spread in convergence behavior arising from different spatial arrangements of blocked angular sectors at each density level. The curves exhibit the typical GS profile: a rapid initial decrease followed by a plateau, with the vertical dashed line indicating the conservative iteration cap $I_{\max}=40$. Notably, larger blockage fractions yield faster stabilization—pruning reduces the enforced angular domain and improves the conditioning of $\mathbf A\mathbf A^{H}$ on the active grid—so fewer LS–projection cycles are required to satisfy the stopping criteria. 


\begin{figure}
\centering
        \includegraphics[width=1.01\linewidth]{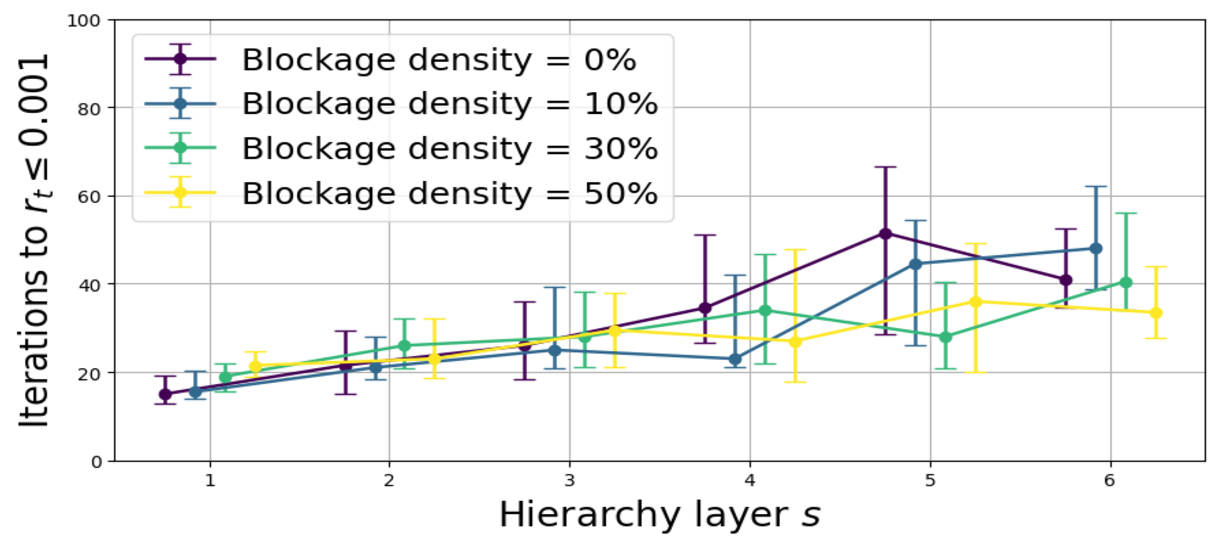}
        \caption{Iterations to stabilization by hierarchy layer and blockage density. For each hierarchy layer $s$, the figure reports the median number of iterations required to reach the convergence threshold $r_t \le r_\star = 10^{-3}$, with vertical bars denoting the IQR.}
        \vspace{-5mm}
\label{cov3}
\end{figure}

Fig. \ref{cov2} provides a complementary distributional perspective by depicting box plots of the normalized residual $r_t$ at selected iteration indices $t \in \{1, 20, 40, 60, 80, 100\}$ for each blockage density corresponding to Fig. \ref{cov1}. The box boundaries represent the IQR across 240 independent sector realizations, while the solid central line marks the median. Three key observations emerge: (i) all configurations initialize at $r_t \approx 1$; (ii) by $t=20$, median residuals stabilize near 0.66 with IQR spanning approximately 0.35–0.85; and (iii) beyond $t=20$, both medians and IQR widths remain virtually unchanged through $t=100$, indicating rapid convergence followed by stable oscillation within an attractor set. This plateau behavior confirms that the algorithm's effective iteration budget is modest ($< 20$ cycles), with subsequent iterations providing negligible additional residual reduction. This distributional stability confirms that the proposed algorithm converges reliably regardless of the specific blockage pattern realization. 
Fig. \ref{cov3} complements this view by reporting, for each hierarchy layer $s$, the iterations needed to reach the target threshold $r_t\!\le\!r_\star$ (median with inter-quartile bars): deeper layers with narrower sectors require slightly more iterations due to tighter beam specifications and increased sensitivity, while increased blockage consistently reduces iteration counts across all layers by constraining the feasible space. Together, the figures justify a single global $I_{\max}$ that upper-bounds the observed stopping iterations across layers and blockage densities, while empirically confirming that blockage-aware pruning accelerates convergence throughout the hierarchy.
\section{Simulation Results}
The proposed scheme is evaluated using the parameter configurations specified in Table \ref{tab1}, which adhere to the mmWave environment parameter settings defined in \cite{berraki2014codebook,lu2023hierarchical}. The RIS structure is a uniform linear array with antenna spacing $d = \lambda/2$, where $\lambda = 0.005$ m, corresponding to the 60 GHz frequency. Based on the analysis presented in \cite{wei2025mechanical}, the XLA series of high-speed miniature linear actuators from XERYON is considered as a candidate for driving the MA \cite{xeryonActuator}. With a maximum velocity of up to 1000 mm/s, it enables the mechanical actuation time of the MA to be constrained to the millisecond scale with millimeter-level positioning precision, thereby we set the time $\tau = 0.5$ ms to meet the stringent temporal requirements imposed by limited training durations. Additionally, the training overhead is concretely quantified as the minimum number of beam evaluations required for each method to achieve a target rate threshold of 80\% of the average optimal rate across all users under perfect CSI, where fewer beam evaluations correspond to reduced training time, since each evaluation requires pilot transmission and channel measurement duration. 

\begin{table}
\caption{System simulation parameters}
\label{table}
\setlength{\tabcolsep}{3pt}
\begin{tabular}{|p{30pt}|p{130pt}|p{50pt}|}
\hline
Symbol& 
Quantity& 
Value\\
\hline
$M$& 
No. of MAs& 
64\\
$N$& 
No. of RIS elements& 
256\\
$K$& 
No. of UEs& 
2\\
$L_g$& 
No. of MA-BS to RIS paths& 
9\\
$L_b$& 
No. of RIS to UE paths& 
5\\
$f_c$& 
Carrier frequency& 
60 GHz\\
$B$& 
Bandwidth& 
1 GHz\\
$\sigma^2$& 
Noise power & 
-90 dBm\\
$\tau$& 
MA mechanical motion time& 
0.5 ms\\
$T$& 
Total training time& 
200 ms\\
\hline
\end{tabular}
\label{tab1}
\end{table}

\subsection{Achievable rate performance comparison}
\begin{figure}[!t]
\centering
\includegraphics[width=1.01\linewidth, trim=0 0 0 25, clip]{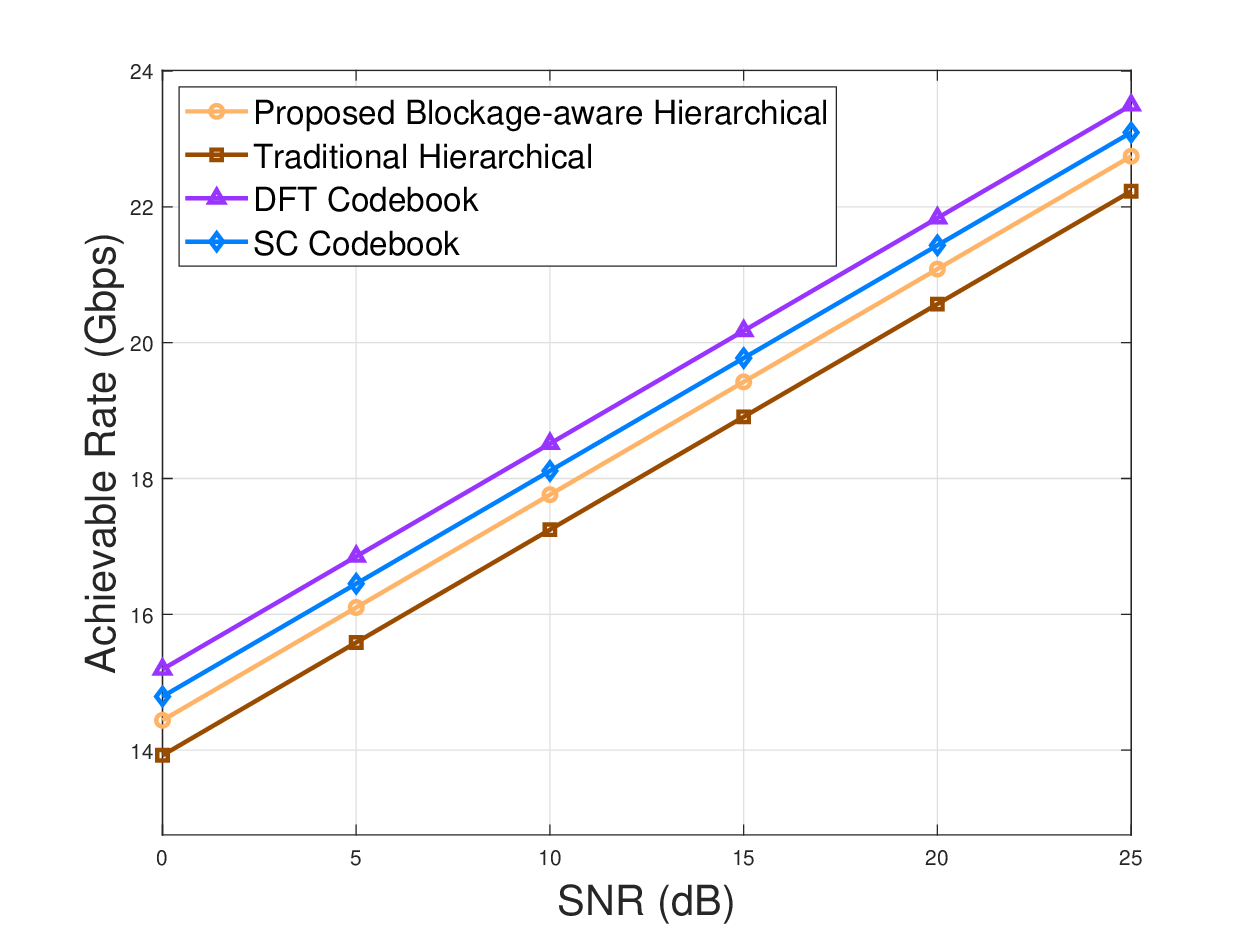}
\caption{Achievable rate versus SNR under different codebook structures. The proposed blockage-aware hierarchical codebook is compared with a traditional hierarchical codebook, a DFT codebook, and a subspace (SC) codebook. Unless otherwise stated, parameters follow Table~I (e.g., $M{=}64$, $N{=}256$, $K{=}2$, $f_c{=}60$\,GHz, $B{=}1$\,GHz, $\sigma^2{=}{-}90$\,dBm).}
\label{rate_snr}
\end{figure}
\begin{figure}[!t]
\centering
\includegraphics[width=1.01\linewidth, trim=0 0 0 25, clip]{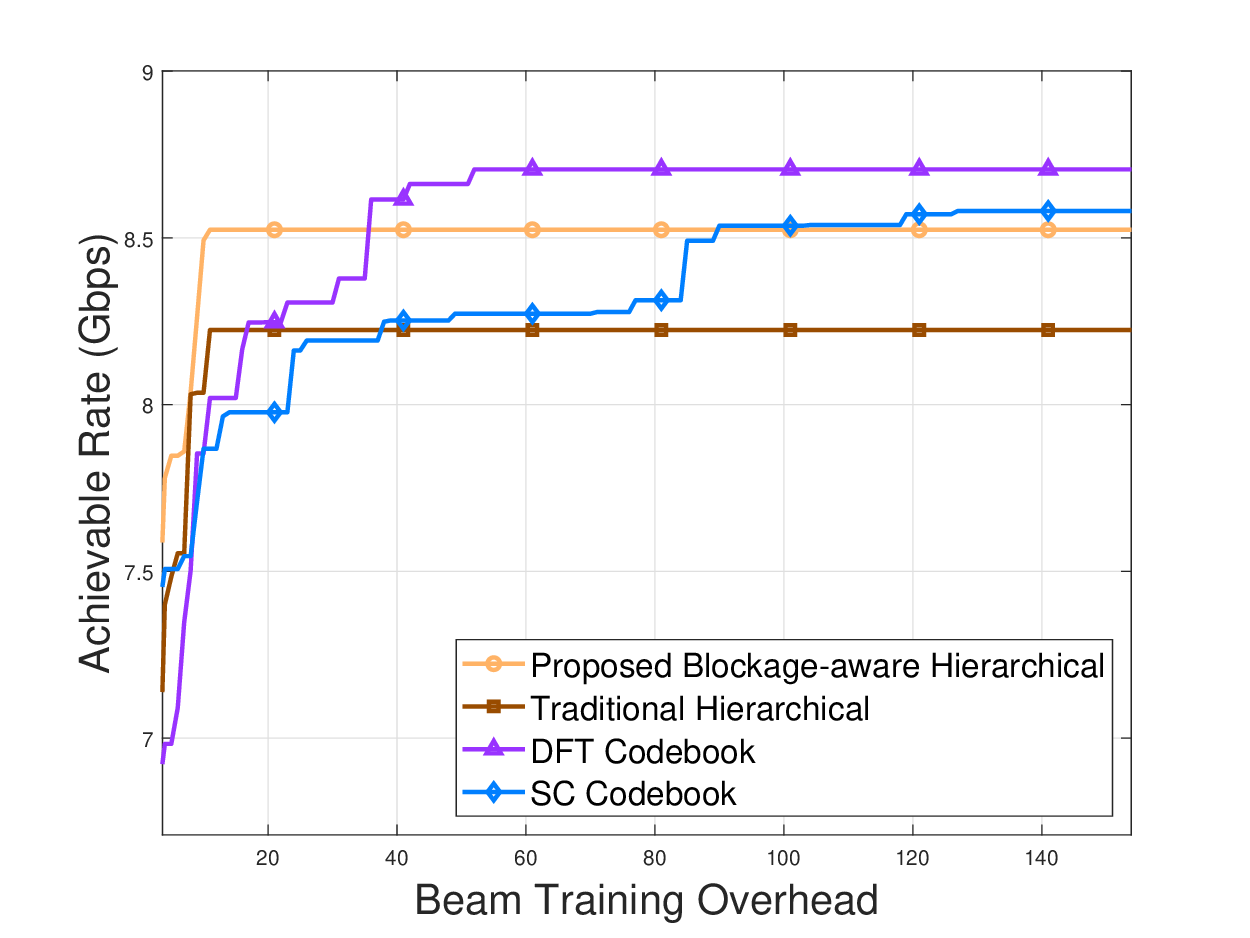}
\caption{Achievable rate versus beam-training overhead (number of beam evaluations). Curves are averaged over channel and blockage realizations with the same settings as in Fig.~\ref{rate_snr}. The proposed blockage-aware hierarchy reaches a stable operating point with markedly fewer evaluations than the traditional hierarchical, DFT, and SC baselines by skipping blocked directions and enforcing a blockage-masked magnitude profile per node via GS.}
\vspace{-5mm}
\label{rate_overhead}
\end{figure}
We conduct simulation of achievable rate vs. SNR and vs. overhead cost separately. The performance evaluation reveals a critical trade-off of achievable rate and training efficiency in blockage-aware design. While in Fig. \ref{rate_snr} we can see that our proposed blockage-aware hierarchical codebook achieves a slightly lower rate than existing schemes, in Fig.~\ref{rate_overhead} we can see that it reaches a stable performance with significantly less training overhead, i.e., our proposed method exhibits a faster convergence toward its maximum rate level as compared to other methods. The DFT \cite{lee2016channel} achieves the highest rate (8.7 Gbps) but requires substantially more beam training evaluations to reach the inflection point. Our method trades a small amount of peak rate (about 0.1 Gbps) compared to Subspace Codebook (SC) \cite{shen2018channel} for a sizable reduction of about $80\%$ in training overhead. The earlier inflection point arises from the hierarchical structure’s pre-selection of unblocked angles, enabling rapid focus on viable beam directions without exhaustive search.
\begin{figure}[!htbp]
\centering
\includegraphics[width=1.01\linewidth, trim=0 0 0 25, clip]{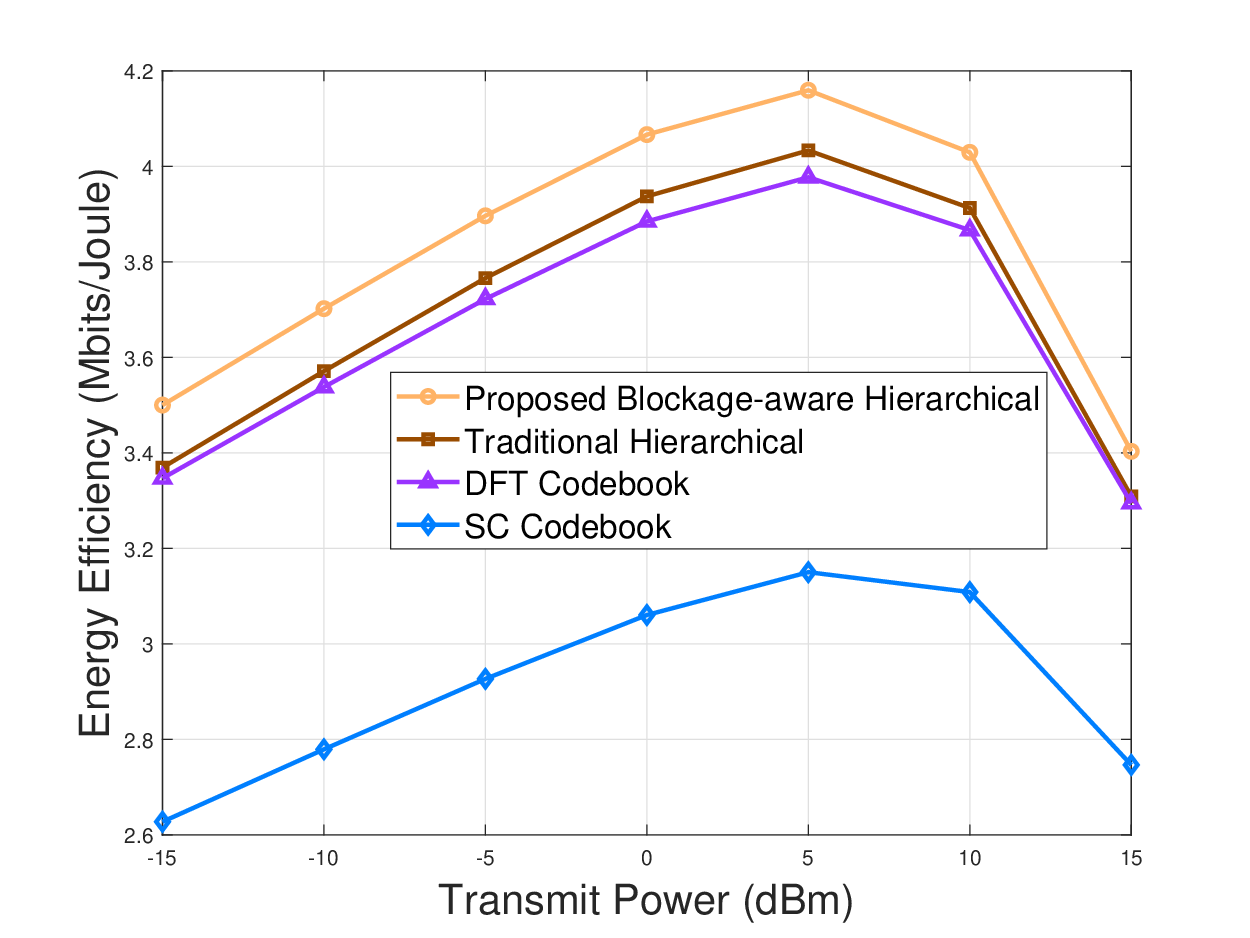}
\caption{Energy efficiency (EE) versus SNR. EE is defined as $\eta_{\mathrm{EE}}=\frac{(T-\tau)C}{\tau P_M + \frac{(T-\tau)P_D}{\eta_{\mathrm{amp}}}}$, with $\tau{=}0.5$\,ms and $T{=}200$\,ms.}
\vspace{-5mm}
\label{EEvsPt}
\end{figure}

\subsection{EE performance comparison}
We set the parameters for the EE model in accordance to \cite{wei2025mechanical}: the static circuit power consumption is $P_{s} = 100$ mW, the dynamic power consumption is $P_{c} = 300$ mW. As the power amplifier efficiency $\eta_{amp}$ typically achieves a value between 0.2 and 0.4 \cite{guo2015power}, we set its value as $\eta_{amp} = 0.2$.
As illustrated in Fig.~\ref{EEvsPt}, our design achieves the highest EE as compared to traditional hierarchical, DFT codebook, and SC codebook methods. This improvement stems from the adaptive beam rotation strategy, which optimizes the use of MA by avoiding blocked regions, thereby reducing unnecessary power consumption. However, the high energy cost associated with MA operation poses a significant challenge. The proposed method mitigates this by leveraging the hierarchical structure to minimize active antenna adjustments, however, as we can see there exists a peak point at approximately 4.2 Mbits/Joule which is associated with moderate transmit power (around 5 dBm), i.e., there exists a trade-off between data rate and power consumption. Our blockage-aware design excels by leveraging this trade-off strategically, accepting a moderate rate reduction to operate closer to the energy efficiency-optimal region, making it particularly valuable for energy-constrained massive MIMO deployments where battery life and thermal management are paramount. 

\subsection{Overhead cost vs. blockage density}
\begin{figure}[!t]
\centering
\includegraphics[width=1.01\linewidth, trim=0 0 0 25, clip]{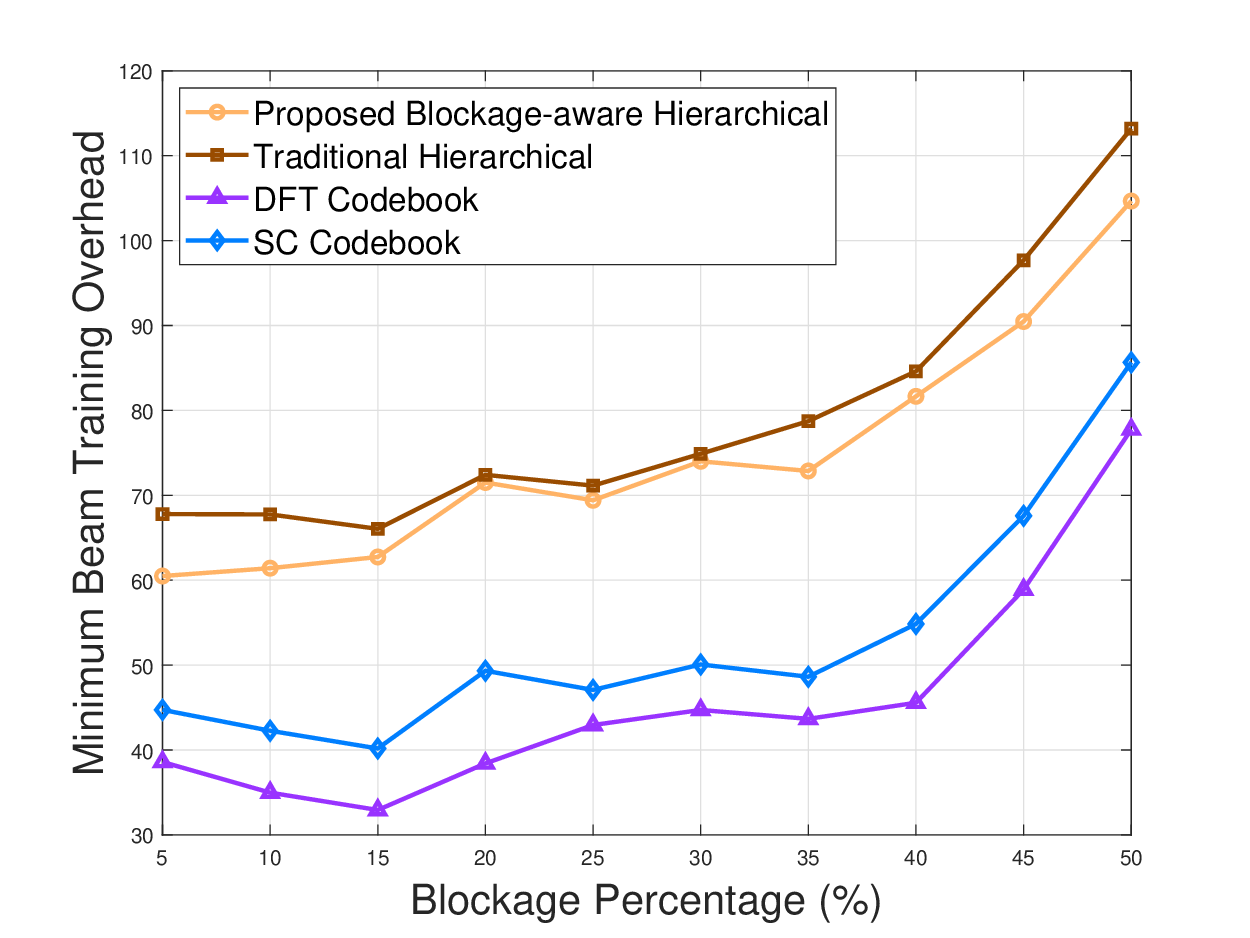}
\caption{Energy efficiency (EE) versus beam-training overhead. Consistent with Fig.~\ref{rate_overhead}, the proposed method attains near-peak EE at significantly lower overhead by pruning blocked branches early in the hierarchy and accelerating GS convergence. Baselines require more evaluations to approach their best EE, increasing pilot/measurement energy and diminishing net efficiency within a coherence interval.}
\vspace{-5mm}
\label{OverheadvsBlo}
\end{figure}

To analyze the sensitivity of each method to varying blockage densities — reflecting more realistic environments — we simulate training overhead across different blockage levels. The different blockage levels represent the proportion of angular space obstructed by randomly placed obstacles. In Fig.~\ref{OverheadvsBlo}, the proposed blockage-aware hierarchical codebook consistently achieves lower overhead compared to the traditional hierarchical approach, demonstrating a reduction in training complexity through blockage-adaptive beam selection. DFT and SC codebooks exhibit minimal overhead at low blockage densities with observable fluctuations attributed to the stochastic nature of dynamic blockage placement, where certain blockage configurations may randomly align or misalign with the predetermined beam patterns.
\section{CONCLUSION}
The proposed blockage-aware hierarchical codebook design offers a compelling trade-off in performance metrics. Its superior EE and reduced beam training overhead make it highly suitable for energy-constrained and dynamic environments, particularly for systems employing MAs in those scenarios. The high energy cost of MA operation, a significant challenge, is effectively managed through the hierarchical and GS-based refinement strategies, which optimize beam coverage without excessive power expenditure. Importantly, the convergence analysis confirms that codebook generation complexity remains tractable: the GS iterations stabilize rapidly (within 40 iterations across all scenarios), and higher blockage densities actually accelerate convergence, making the approach particularly efficient in the challenging environments where it is most needed. However, this comes at the cost of a reduced achievable rate versus SNR, a trade-off deemed acceptable given the gains in EE and overhead reduction. This balanced approach enhances the practicality of large-scale MIMO systems in blockage-prone near-field scenarios. Future work will explore advanced joint optimization techniques that simultaneously optimize MA-BS antenna positioning and RIS phase shifts to improve the rate performance of our method to match the best-performing schemes, while maintaining its favourable EE and training overhead characteristics.

\normalem
\bibliographystyle{unsrt}
\bibliography{reference}

@article{karacora2025robust,
  title={Robust Communication Design in RIS-Assisted THz Channels},
  author={Karacora, Yasemin and Umra, Adam and Sezgin, Aydin},
  journal={IEEE Open Journal of the Communications Society},
  year={2025},
  publisher={IEEE}
}

@article{wang2024tutorial,
  title={A tutorial on extremely large-scale MIMO for 6G: Fundamentals, signal processing, and applications},
  author={Wang, Zhe and Zhang, Jiayi and Du, Hongyang and Niyato, Dusit and Cui, Shuguang and Ai, Bo and Debbah, M{\'e}rouane and Letaief, Khaled B and Poor, H Vincent},
  journal={IEEE Communications Surveys \& Tutorials},
  volume={26},
  number={3},
  pages={1560--1605},
  year={2024},
  publisher={IEEE}
}

@article{zhu2023modeling,
  title={{Modeling and performance analysis for movable antenna enabled wireless communications}},
  author={Zhu, Lipeng and Ma, Wenyan and Zhang, Rui},
  journal={IEEE Transactions on Wireless Communications},
  volume={23},
  number={6},
  pages={6234--6250},
  year={2023},
  publisher={IEEE}
}

@article{shao2025tutorial,
  title={A tutorial on six-dimensional movable antenna for 6G networks: Synergizing positionable and rotatable antennas},
  author={Shao, Xiaodan and Mei, Weidong and You, Changsheng and Wu, Qingqing and Zheng, Beixiong and Wang, Cheng-Xiang and Li, Junling and Zhang, Rui and Schober, Robert and Zhu, Lipeng and others},
  journal={arXiv preprint arXiv:2503.18240},
  year={2025}
}

@article{ning2024movable,
  title={{Movable antenna-enhanced wireless communications: General architectures and implementation methods}},
  author={Ning, Boyu and Yang, Songjie and Wu, Yafei and Wang, Peilan and Mei, Weidong and Yuen, Chau and Bj{\"o}rnson, Emil},
  journal={arXiv preprint arXiv:2407.15448},
  year={2024}
}

@inproceedings{bai2013coverage,
  title={{Coverage analysis for millimeter wave cellular networks with blockage effects}},
  author={Bai, Tianyang and Heath, Robert W},
  booktitle={2013 IEEE Global Conference on Signal and Information Processing},
  pages={727--730},
  year={2013},
  organization={IEEE}
}

@article{palitharathna2023neural,
  title={{Neural network-based blockage prediction and optimization in lightwave power transfer-enabled hybrid VLC/RF systems}},
  author={Palitharathna, Kapila WS and Suraweera, Himal A and Godaliyadda, Roshan I and Herath, Vijitha R and Ding, Zhiguo},
  journal={IEEE Internet of Things Journal},
  year={2023},
  publisher={IEEE}
}

@article{lu2023hierarchical,
  title={{Hierarchical beam training for extremely large-scale MIMO: From far-field to near-field}},
  author={Lu, Yu and Zhang, Zijian and Dai, Linglong},
  journal={IEEE Transactions on Communications},
  volume={72},
  number={4},
  pages={2247--2259},
  year={2023},
  publisher={IEEE}
}

@inproceedings{hwang2024movable,
  title={{Movable-Antenna-Array-Aided Hierarchical Codebook Design}},
  author={Hwang, Seungjin and Park, Juhyun and Lee, Chungyong},
  booktitle={2024 IEEE International Conference on Consumer Electronics-Asia (ICCE-Asia)},
  pages={1--5},
  year={2024},
  organization={IEEE}
}

@article{zhu2024movable,
  author={Zhu, Lipeng and Ma, Wenyan and Zhang, Rui},
  title={Movable Antennas for Wireless Communication: Opportunities and Challenges},
  journal={IEEE Communications Magazine},
  year={2024},
  volume={62},
  number={6},
  pages={114--120},
  doi={10.1109/MCOM.001.2300212}
}

@article{ma2025robust,
  title={Robust Movable-Antenna Position Optimization with Imperfect CSI for MISO Systems},
  author={Ma, Haifeng and Mei, Weidong and Wei, Xin and Ning, Boyu and Chen, Zhi},
  journal={IEEE Communications Letters},
  year={2025},
  publisher={IEEE}
}

@article{zhu2023movable,
  title={Movable-antenna enhanced multiuser communication via antenna position optimization},
  author={Zhu, Lipeng and Ma, Wenyan and Ning, Boyu and Zhang, Rui},
  journal={IEEE Transactions on Wireless Communications},
  volume={23},
  number={7},
  pages={7214--7229},
  year={2023},
  publisher={IEEE}
}

@article{yan2025movable,
  title={Movable antenna aided multiuser communications: Antenna position optimization based on statistical channel information},
  author={Yan, Ge and Zhu, Lipeng and Zhang, Rui},
  journal={arXiv preprint arXiv:2502.20856},
  year={2025}
}

@inproceedings{chen2023joint,
  title={Joint beamforming and antenna movement design for moveable antenna systems based on statistical CSI},
  author={Chen, Xintai and Feng, Biqian and Wu, Yongpeng and Ng, Derrick Wing Kwan and Schober, Robert},
  booktitle={GLOBECOM 2023-2023 IEEE Global Communications Conference},
  pages={4387--4392},
  year={2023},
  organization={IEEE}
}

@article{zeng2024csi,
  title={CSI-Free Position Optimization for Movable Antenna Communication Systems: A Black-Box Optimization Approach},
  author={Zeng, Xianlong and Fang, Jun and Wang, Bin and Ning, Boyu and Li, Hongbin},
  journal={IEEE Wireless Communications Letters},
  year={2024},
  publisher={IEEE}
}

@article{zeng2025derivative,
  title={A Derivative-Free Position Optimization Approach for Movable Antenna Multi-User Communication Systems},
  author={Zeng, Xianlong and Fang, Jun and Wang, Peilan and Mei, Weidong and Liang, Ying-Chang},
  journal={arXiv preprint arXiv:2505.19012},
  year={2025}
}

@article{ma2023mimo,
  title={MIMO capacity characterization for movable antenna systems},
  author={Ma, Wenyan and Zhu, Lipeng and Zhang, Rui},
  journal={IEEE Transactions on Wireless Communications},
  volume={23},
  number={4},
  pages={3392--3407},
  year={2023},
  publisher={IEEE}
}

@article{ma2024movable,
  title={Movable antenna enhanced wireless sensing via antenna position optimization},
  author={Ma, Wenyan and Zhu, Lipeng and Zhang, Rui},
  journal={IEEE Transactions on Wireless Communications},
  year={2024},
  publisher={IEEE}
}

@article{brilhante2023literature,
  title={A literature survey on AI-aided beamforming and beam management for 5G and 6G systems},
  author={Brilhante, Davi da Silva and Manjarres, Joanna Carolina and Moreira, Rodrigo and de Oliveira Veiga, Lucas and de Rezende, Jos{\'e} F and M{\"u}ller, Francisco and Klautau, Aldebaro and Leonel Mendes, Luciano and P. de Figueiredo, Felipe A},
  journal={Sensors},
  volume={23},
  number={9},
  pages={4359},
  year={2023},
  publisher={MDPI}
}

@article{eom2025hierarchical,
  title={Hierarchical Codebook Design for Beam Alignment in Terahertz Ultra-Massive MIMO Systems},
  author={Eom, Chahyeon and Lee, Chungyong},
  journal={IEEE Internet of Things Journal},
  year={2025},
  publisher={IEEE}
}

@article{xiao2016hierarchical,
  author={Xiao, Zhenyu and others},
  title={Hierarchical Codebook Design for Beamforming Training in Millimeter-Wave Communication},
  journal={IEEE Transactions on Wireless Communications},
  year={2016},
  volume={15},
  number={5},
  pages={3380--3392},
  doi={10.1109/TWC.2016.2519401}
}

@article{wang2009beam,
  title={Beam codebook based beamforming protocol for multi-Gbps millimeter-wave WPAN systems},
  author={Wang, Junyi and Lan, Zhou and Pyo, Chang-woo and Baykas, Tuncer and Sum, Chin-sean and Rahman, Mohammad Azizur and Gao, Jing and Funada, Ryuhei and Kojima, Fumihide and Harada, Hiroshi and others},
  journal={IEEE Journal on Selected Areas in Communications},
  volume={27},
  number={8},
  pages={1390--1399},
  year={2009},
  publisher={IEEE}
}

@article{matsumura2011orthogonal,
  title={Orthogonal beamforming using Gram-Schmidt orthogonalization for multi-user MIMO downlink system},
  author={Matsumura, Kunitaka and Ohtsuki, Tomoaki},
  journal={Eurasip journal on wireless communications and networking},
  volume={2011},
  number={1},
  pages={41},
  year={2011},
  publisher={Springer}
}

@article{wang2023hierarchical,
  title={Hierarchical codebook-based beam training for RIS-assisted mmWave communication systems},
  author={Wang, Jinghe and Tang, Wankai and Jin, Shi and Wen, Chao-Kai and Li, Xiao and Hou, Xiaolin},
  journal={IEEE Transactions on Communications},
  volume={71},
  number={6},
  pages={3650--3662},
  year={2023},
  publisher={IEEE}
}

@article{shi2023spatial,
  title={Spatial-chirp codebook-based hierarchical beam training for extremely large-scale massive MIMO},
  author={Shi, Xu and Wang, Jintao and Sun, Zhi and Song, Jian},
  journal={IEEE Transactions on Wireless Communications},
  volume={23},
  number={4},
  pages={2824--2838},
  year={2023},
  publisher={IEEE}
}

@article{liu2022adaptive,
  title={Adaptive beam search for initial beam alignment in millimetre-wave communications},
  author={Liu, Chunshan and Zhao, Lou and Li, Min and Yang, Lei},
  journal={IEEE Transactions on Vehicular Technology},
  volume={71},
  number={6},
  pages={6801--6806},
  year={2022},
  publisher={IEEE}
}

@ARTICLE{6840343,
  author={Bai, Tianyang and Vaze, Rahul and Heath, Robert W.},
  journal={IEEE Transactions on Wireless Communications}, 
  title={Analysis of Blockage Effects on Urban Cellular Networks}, 
  year={2014},
  volume={13},
  number={9},
  pages={5070-5083},
  doi={10.1109/TWC.2014.2331971}}

@article{ramirez2024observations,
  title={Observations on large-scale attenuation effects in a 26 GHz urban micro-cell environment},
  author={Ram{\'\i}rez-Arroyo, Alejandro and S{\o}rensen, Troels B and Beltoft, Peter and Christiansen, Henrik and Valenzuela-Vald{\'e}s, Juan F and Mogensen, Preben},
  journal={IEEE Wireless Communications Letters},
  volume={13},
  number={9},
  pages={2611--2615},
  year={2024},
  publisher={IEEE}
}

@inproceedings{alyosef2022survey,
  title={A survey on the effects of human blockage on the performance of mm wave communication systems},
  author={Alyosef, Ayham and Rizou, Stamatia and Zaharis, Zaharias D and Lazaridis, Pavlos I and Nor, Ahmed M and Fratu, Octavian and Halunga, Simona and Yioultsis, Traianos V and Kantartzis, Nikolaos V},
  booktitle={2022 IEEE International Black Sea Conference on Communications and Networking (BlackSeaCom)},
  pages={249--253},
  year={2022},
  organization={IEEE}
}

@article{azpilicueta2020fifth,
  title={Fifth-generation (5G) mmwave spatial channel characterization for urban environments’ system analysis},
  author={Azpilicueta, Leyre and Lopez-Iturri, Peio and Zu{\~n}iga-Mejia, Jaime and Celaya-Echarri, Mikel and Rodr{\'\i}guez-Corbo, Fidel Alejandro and Vargas-Rosales, Cesar and Aguirre, Erik and Michelson, David G and Falcone, Francisco},
  journal={Sensors},
  volume={20},
  number={18},
  pages={5360},
  year={2020},
  publisher={MDPI}
}

@inproceedings{gapeyenko2016analysis,
  title={Analysis of human-body blockage in urban millimeter-wave cellular communications},
  author={Gapeyenko, Margarita and Samuylov, Andrey and Gerasimenko, Mikhail and Moltchanov, Dmitri and Singh, Sarabjot and Aryafar, Ehsan and Yeh, Shu-ping and Himayat, Nageen and Andreev, Sergey and Koucheryavy, Yevgeni},
  booktitle={2016 IEEE International Conference on Communications (ICC)},
  pages={1--7},
  year={2016},
  organization={IEEE}
}

@article{liu2021blockage,
  title={Blockage tolerance in roadside millimeter-wave backhaul networks},
  author={Liu, Yuchen and Blough, Douglas M},
  journal={Computer Networks},
  volume={198},
  pages={108377},
  year={2021},
  publisher={Elsevier}
}

@article{wu2025intelligent,
  title={Intelligent reflecting surfaces for wireless networks: Deployment architectures, key solutions, and field trials},
  author={Wu, Qingqing and Chen, Guangji and Peng, Qiaoyan and Chen, Wen and Yuan, Yifei and Cheng, Zhenqiao and Dou, Jianwu and Zhao, Zhiyong and Li, Ping},
  journal={IEEE Wireless Communications},
  year={2025},
  publisher={IEEE}
}

@article{geng2025joint,
  title={Joint Beamforming and Antenna Position Optimization for IRS-Aided Multi-User Movable Antenna Systems},
  author={Geng, Yue and Cheng, Tee Hiang and Zhong, Kai and Teh, Kah Chan and Wu, Qingqing},
  journal={IEEE Transactions on Wireless Communications},
  year={2025},
  publisher={IEEE}
}

@article{yu2025achievable,
  title={Achievable Rate Optimization for Reconfigurable Intelligent Surface-Aided Multi-User Movable Antenna Systems},
  author={Yu, Liji and Ren, Yuhui},
  journal={Sensors},
  volume={25},
  number={15},
  pages={4694},
  year={2025},
  publisher={MDPI}
}

@article{yang2025robust,
  title={Robust Transceiver Design for RIS Enhanced Dual-Functional Radar-Communication with Movable Antenna},
  author={Yang, Ran and Dong, Zheng and Cheng, Peng and Lyu, Wanting and Xiu, Yue and Li, Yan and Wei, Ning},
  journal={arXiv preprint arXiv:2506.07610},
  year={2025}
}

@article{callebaut2021art,
  title={The art of designing remote iot devices—technologies and strategies for a long battery life},
  author={Callebaut, Gilles and Leenders, Guus and Van Mulders, Jarne and Ottoy, Geoffrey and De Strycker, Lieven and Van der Perre, Liesbet},
  journal={Sensors},
  volume={21},
  number={3},
  pages={913},
  year={2021},
  publisher={MDPI}
}

@article{rzig2025energy,
  title={Energy-Efficient Vehicular Task Offloading Using Multi-Mode MEC and RIS-Equipped Aerial Platforms},
  author={Rzig, Insaf and Jaafar, Wael and Jebalia, Maha and Tabbane, Sami},
  journal={IEEE Open Journal of the Communications Society},
  year={2025},
  publisher={IEEE}
}

@article{zhao2024joint,
  title={Joint Power Allocation and Beamforming Design for Active IRS-Aided Secure Directional Modulation Systems},
  author={Zhao, Yifan and Wang, Xiaoyu and Zhou, Kaibo and Wang, Xuehui and Wang, Yan and Gao, Wei and Liu, Ruiqi and Shu, Feng},
  journal={IEEE Open Journal of the Communications Society},
  year={2024},
  publisher={IEEE}
}

@article{tang2021battery,
  title={Battery-constrained federated edge learning in UAV-enabled IoT for B5G/6G networks},
  author={Tang, Shunpu and Zhou, Wenqi and Chen, Lunyuan and Lai, Lijia and Xia, Junjuan and Fan, Liseng},
  journal={Physical Communication},
  volume={47},
  pages={101381},
  year={2021},
  publisher={Elsevier}
}

@article{wu2019intelligent,
  title={Intelligent reflecting surface enhanced wireless network via joint active and passive beamforming},
  author={Wu, Qingqing and Zhang, Rui},
  journal={IEEE transactions on wireless communications},
  volume={18},
  number={11},
  pages={5394--5409},
  year={2019},
  publisher={IEEE}
}

@article{di2020smart,
  title={Smart radio environments empowered by reconfigurable intelligent surfaces: How it works, state of research, and the road ahead},
  author={Di Renzo, Marco and Zappone, Alessio and Debbah, Merouane and Alouini, Mohamed-Slim and Yuen, Chau and De Rosny, Julien and Tretyakov, Sergei},
  journal={IEEE journal on selected areas in communications},
  volume={38},
  number={11},
  pages={2450--2525},
  year={2020},
  publisher={IEEE}
}

@article{mukherjee2022leveraging,
  title={Leveraging big data analytics in 5G-enabled IoT and industrial IoT for the development of sustainable smart cities},
  author={Mukherjee, Suprakash and Gupta, Shashank and Rawlley, Oshin and Jain, Siddhant},
  journal={Transactions on Emerging Telecommunications Technologies},
  volume={33},
  number={12},
  pages={e4618},
  year={2022},
  publisher={Wiley Online Library}
}

@article{perera2013context,
  title={Context aware computing for the internet of things: A survey},
  author={Perera, Charith and Zaslavsky, Arkady and Christen, Peter and Georgakopoulos, Dimitrios},
  journal={IEEE communications surveys \& tutorials},
  volume={16},
  number={1},
  pages={414--454},
  year={2013},
  publisher={Ieee}
}

@article{shaikh2016energy,
  title={Energy harvesting in wireless sensor networks: A comprehensive review},
  author={Shaikh, Faisal Karim and Zeadally, Sherali},
  journal={Renewable and Sustainable Energy Reviews},
  volume={55},
  pages={1041--1054},
  year={2016},
  publisher={Elsevier}
}

@article{bjornson2017massive,
  title={Massive MIMO networks: Spectral, energy, and hardware efficiency},
  author={Bj{\"o}rnson, Emil and Hoydis, Jakob and Sanguinetti, Luca and others},
  journal={Foundations and Trends{\textregistered} in Signal Processing},
  volume={11},
  number={3-4},
  pages={154--655},
  year={2017},
  publisher={Now Publishers, Inc.}
}

@article{wang2020channel,
  title={{Channel estimation for intelligent reflecting surface assisted multiuser communications: Framework, algorithms, and analysis}},
  author={Wang, Zhaorui and Liu, Liang and Cui, Shuguang},
  journal={IEEE Transactions on Wireless Communications},
  volume={19},
  number={10},
  pages={6607--6620},
  year={2020},
  publisher={IEEE}
}

@article{galiotto2017effect,
  title={{Effect of LOS/NLOS propagation on 5G ultra-dense networks}},
  author={Galiotto, Carlo and Pratas, Nuno K and Doyle, Linda and Marchetti, Nicola},
  journal={Computer Networks},
  volume={120},
  pages={126--140},
  year={2017},
  publisher={Elsevier}
}

@inproceedings{guo2015power,
  title={{Power allocation for massive MIMO: impact of power amplifier efficiency}},
  author={Guo, Yingchu and Tang, Junlin and Wu, Gang and Li, Shaoqian},
  booktitle={2015 IEEE/CIC International Conference on Communications in China (ICCC)},
  pages={1--6},
  year={2015},
  organization={IEEE}
}

@article{wei2025mechanical,
  title={{Mechanical Power Modeling and Energy Efficiency Maximization for Movable Antenna Systems}},
  author={Wei, Xin and Mei, Weidong and Huang, Xuan and Chen, Zhi and Ning, Boyu},
  journal={arXiv preprint arXiv:2505.05914},
  year={2025}
}

@article{chen2023hierarchical,
  title={{Hierarchical codebook design for near-field mmwave MIMO communications systems}},
  author={Chen, Jiawei and Gao, Feifei and Jian, Mengnan and Yuan, Wanmai},
  journal={{IEEE Wireless Communications Letters}},
  year={2023},
  publisher={IEEE}
}

@article{qi2020hierarchical,
  title={{Hierarchical codebook-based multiuser beam training for millimeter wave massive MIMO}},
  author={Qi, Chenhao and Chen, Kangjian and Dobre, Octavia A and Li, Geoffrey Ye},
  journal={IEEE Transactions on Wireless Communications},
  volume={19},
  number={12},
  pages={8142--8152},
  year={2020},
  publisher={IEEE}
}

@inproceedings{berraki2014codebook,
  title={{Codebook based beamforming and multiuser scheduling scheme for mmWave outdoor cellular systems in the 28, 38 and 60GHz bands}},
  author={Berraki, Djamal E and Armour, Simon MD and Nix, Andrew R},
  booktitle={2014 IEEE Globecom Workshops (GC Wkshps)},
  pages={382--387},
  year={2014},
  organization={IEEE}
}

@online{xeryonActuator,
  author       = {{Xeryon}},
  title        = {{Mini Linear Actuators}},
  year         = {n.d.},
  url          = {https://xeryon.com/products/mini-linear-actuators/},
  note         = {[Online]. Available: \url{https://xeryon.com/products/mini-linear-actuators/}}
}

@article{lee2016channel,
  title={Channel estimation via orthogonal matching pursuit for hybrid MIMO systems in millimeter wave communications},
  author={Lee, Junho and Gil, Gye-Tae and Lee, Yong H},
  journal={IEEE Transactions on Communications},
  volume={64},
  number={6},
  pages={2370--2386},
  year={2016},
  publisher={IEEE}
}

@article{shen2018channel,
  title={{Channel feedback based on AoD-adaptive subspace codebook in FDD massive MIMO systems}},
  author={Shen, Wenqian and Dai, Linglong and Shim, Byonghyo and Wang, Zhaocheng and Heath, Robert W},
  journal={IEEE Transactions on Communications},
  volume={66},
  number={11},
  pages={5235--5248},
  year={2018},
  publisher={IEEE}
}

\end{document}